**Interplay between musical practices and tuning in the *marimba de chonta* music**

Jorge E. Useche[1], Rafael G. Hurtado[1,*] and Federico Demmer[2]

## Abstract

In the Pacific Coast of Colombia there is a type of marimba called *marimba de chonta*, which provides the melodic and harmonic contour for traditional music with characteristic chants and dances. The tunings of this marimba are based on the voice of female singers and allows musical practices, as a transposition that preserves relative distances between bars. Here we show that traditional tunings are consistent with isotonic scales, and that they have changed in the last three decades due to the influence of Western music. Specifically, low octaves have changed into just octaves. Additionally, consonance properties of this instrument include the occurrence of a broad minimum of dissonance that is used in the musical practices, while the narrow local peaks of dissonance are avoided. We found that the main reason for this is the occurrence of uncertainties in the tunings with respect to the mathematical successions of isotonic scales. We conclude that in this music the emergence of tunings and musical practices cannot be considered as separate issues. Consonance, timbre, and musical practices are entangled.

## Introduction

In the Pacific Coast of Colombia and Ecuador there is a marimba with bars made of the timber from a palm called Chonta (*Bactris jauari*), and tubular resonators made of a Bambuseae called Guadua (*Guadua angustifolia*). This marimba is called *marimba de chonta* and provides the melodic and harmonic contour for a traditional music of African descent. “The marimba music, traditional chants and dances from the Colombia South Pacific region and Esmeraldas Province of Ecuador” were inscribed by UNESCO in the list of Intangible Cultural Heritage of Humanity (Ministerio de Cultura de Colombia, 2010a, 2010b; Instituto Nacional de Patrimonio Cultural de Ecuador, 2014; UNESCO, 2015).

Some instrument makers of the *marimba de chonta* use ancestral techniques for empirically tuning the instrument, resulting in tunings that do not conform to Western musical scales (Ministerio de Cultura de Colombia, 2010a; Miñana 2010a, 2010b). These tunings and the musical practices associated with this marimba remain largely unknown, and they are currently at risk of disappearing (Miñana, 2010b; Ministerio de Cultura de Colombia, 2010a). In the Pacific coast of Colombia, the *marimba de chonta* with a traditional tuning is called “traditional marimba”.

Traditional marimbas are played by one or two musicians, each using two percussion mallets that, commonly, simultaneously hit two different bars (Miñana, 2010; Duque, Sanchez & Tascón, 2009; Hernández, 2007; Ramón, 2001). There are reports of the existence of several tunings, each related to the voices of female singers from a particular territory (Miñana, 2010a, 2010b; Duque, Sanchez & Tascón, 2009); these alternate tunings do not, necessarily, follow a specific mathematical progression conserving musical intervals. The harmonic structure of this music can be summarized in geometrical schemes that contain rules to play two bars simultaneously (harmonic interval) over a base that usually contains seven pitches repeated periodically— in a similar way to the octaves in a diatonic scale. In the case of seven pitches, there are two possible ways for grouping them: by discarding one bar, leading to a hexatonic scale; or discarding two bars, producing a pentatonic scale (Duque, Sanchez & Tascón, 2009; Miñana, 2010a, 2010b). Figure 1 shows the hexatonic and pentatonic scales: white bars are the discarded ones, and bars identified with grey or black colors can be played simultaneously when they have the same color. The largest harmonic interval used has six bars between the two sounded bars (Duque, Sanchez & Tascón, 2009). The schemes shown in Figure 1 can start at any bar, depending on the preference of the female singer, generating a transposition principle that keeps the same relative geometrical distances without conserving the same musical intervals (Duque, Sanchez & Tascón, 2009; Tascón, 2008; Miñana, 2010a, 2010b). The melodic motion in this music uses neighboring bars, usually adjacent ones (Miñana, 2010a), generating soft melodic contours (Patel, 2008).

[1] Department of Physics, Universidad Nacional de Colombia.
[2] Conservatory of Music, Universidad Nacional de Colombia.
[*] Corresponding author. Electronic Mail: rghurtadoh@unal.edu.co

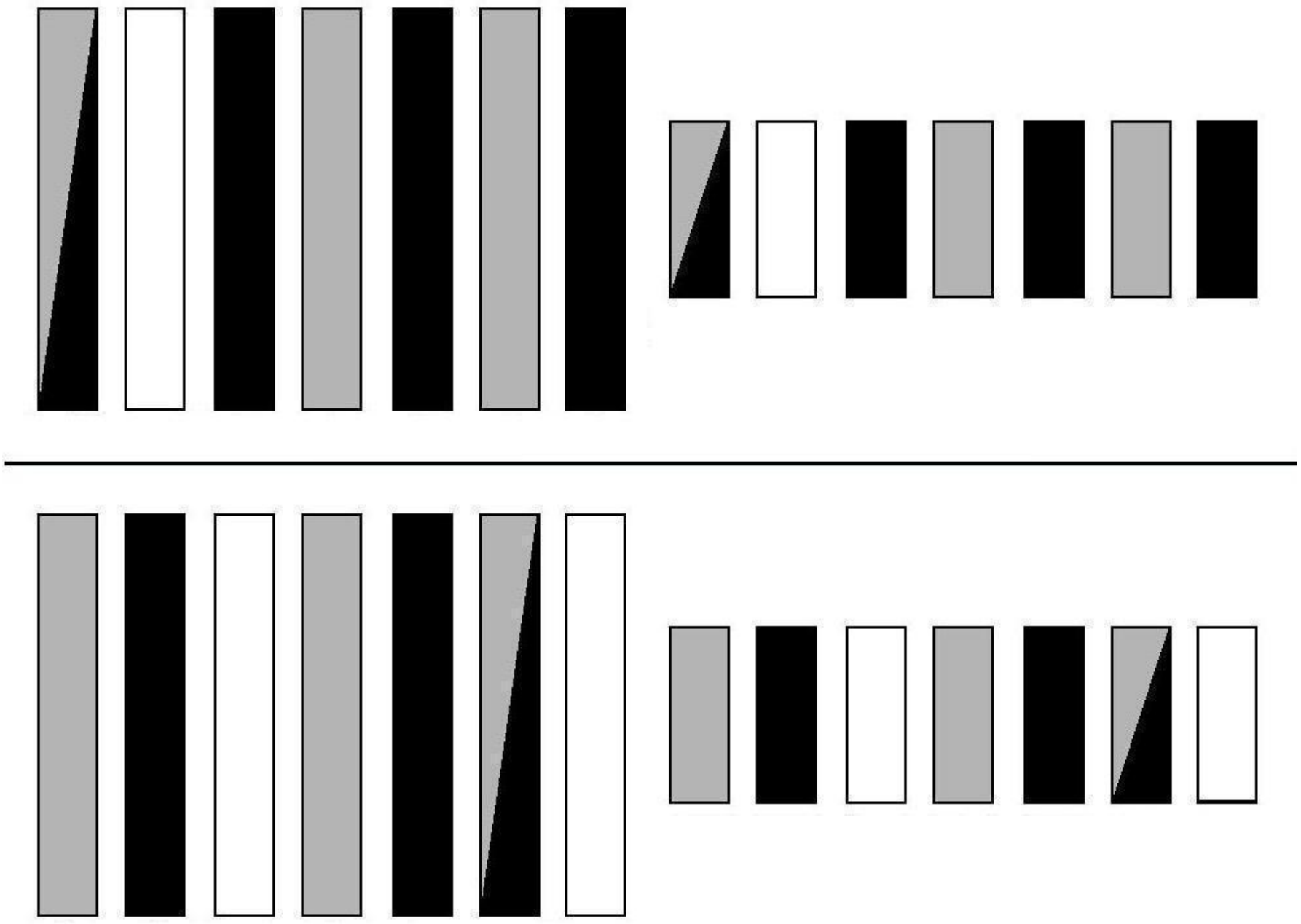

**Figure 1.** Hexatonic (up) and Pentatonic (down) scales generated by an equi-heptatonic traditional marimba. Large bars and small bars indicate two different groups of seven bars. White bars are not used in each scale. Grey and black bars constitute two different harmonic families.

One of the most accepted approaches to the understanding of musical tunings and scales is based on the concept of consonance. For several musical tunings the set of sounds is chosen in such a way as to yield a large number of consonant combinations when two or more elements are sounded together, as, for example, in the just scale (Roederer, 2009). The perception of consonance has been formally related with the sensation of pleasantness (Plomp & Levelt, 1965). Pythagoras posted the existence of this phenomenon when he found that two sounds emitted simultaneously by vibrating strings of equal tension and density produce a pleasant sensation when the ratio between their lengths is the same as the ratio between two small natural numbers. His postulate has been fundamental for parameterizing musical features using physical quantities and for constructing many musical scales, as the just scale (Rossing, 1990; Roederer, 2009).

Hermann von Helmholtz found that the degree of consonance of pairs of pure tones is related to the beats or shocks produced by fluctuations in the peak intensity occurring at the frequency difference of the pure tunes (Helmholtz, 1954). Then, Reinier Plomp and Willem Levelt (1965) reported that the transition range between consonance and dissonance is related with a critical bandwidth that depends on the frequency difference of the corresponding pure tones. This approach to consonance is known as tonal or sensory because it is based on the physical properties of the stimulus, independent of cultural conventions (Schellenberg & Trainor, 1996).

William Sethares (1993) proposed a mathematical function to describe the empirical results obtained by Plomp and Levelt, and used the frequency spectrum for assigning a level of dissonance (instead of consonance) to the timbre of a musical instrument. Sethares posted that the best acoustical tuning of a musical instrument must be inferred from the local minima of dissonance (Sethares, 1993, 1998, 2005). However, in some cases there are musical practices incompatible with tunings constructed in this way. An example is the just tuning resulting from the harmonic timbre of instruments as those made of strings and pipes (Sethares, 1993); this tuning does not divide the octave in equal semitones to produce an equivalent system in all keys (Apel, 1974). The Twelve Tone Equal-Tempered (TTET) scale solves this problem by shifting the frequency ratios of the Pythagoras rule scarifying the high levels of consonance of the just tuning (Apel, 1974).

In this manuscript, in order to study the interplay between consonance, tuning, and musical practices, the Methods section presents the theoretical and experimental procedures to investigate the consonance properties of the *marimba de chonta* and the use of harmonic intervals in its music. The Results and Discussion section contains the musical practices, the tunings of TTET and traditional marimbas de chonta, the consonance properties associated with the timbre of the musical instrument, and the connection between tuning and musical practices. Conclusions are presented in the final section.

## Methods

This section contains the theoretical and experimental procedures used to produce the dissonance level curves for the *marimba de chonta*, make the statistical analysis for the musical pieces, and compare the results obtained in this study with those from a previous one.

### Theoretical

**Dissonance level of individual complex tones.** For the superposition of two pure tones of frequencies $f_1$ and $f_2$, with amplitudes $a_1$ and $a_2$, respectively, Sethares proposed a first model to measure the level of dissonance by parameterizing the Plomp and Levelt empirical results using a mathematical function given by (Sethares, 1993, 1998)

$$d(f_1, f_2, a_1, a_2) = a_{max} \cdot a_{min}\left[e^{-b_1 s(f_{max}-f_{min})} - e^{-b_2 s(f_{max}-f_{min})}\right], \quad (1)$$

where $a_{max} = max(a_1, a_2)$, $a_{min} = min(a_1, a_2)$, $f_{max} = max(f_1, f_2)$, $f_{min} = min(f_1, f_2)$, $b_1 = 3.5$, $b_2 = 5.75$, and $s = 0.24/(0.0207 f_1 + 18.96)$.

Vassilakis modified this model by including the dependence of roughness on intensity, amplitude fluctuation degree, and amplitude fluctuation rate (Vassilakis, 2001; Vassilakis & Kendall, 2010)

$$d(f_1, f_2, a_1, a_2) = (0.5)(a_{max} \cdot a_{min})^{0.1}\left[\frac{2a_{min}}{a_{min} + a_{max}}\right]^{3.11}\left[e^{-b_1 s(f_{max}-f_{min})} - e^{-b_2 s(f_{max}-f_{min})}\right]. \quad (2)$$

Sethares proposed a new model taking into account that the amplitude of the beating is given by the minimum of the two amplitudes, and that the loudness of the roughness is proportional to the loudness of the beating (Sethares, 2005)

$$d(f_1, f_2, a_1, a_2) = \ell_{min}\left[e^{-b_1 s(f_{max}-f_{min})} - e^{-b_2 s(f_{max}-f_{min})}\right], \quad (3)$$

where $\ell_{min}$ refers to the minimum loudness between the loudnesses produced by the two pure tones with frequencies $f_1$ and $f_2$. The loudness $\ell$ can be measured using different models, and is mainly determined by the amplitude. A simple model to approximate loudness is given by the Stevens law (Marks & Florentine, 2010): for a sinusoidal wave of amplitude $a$, the loudness (in *sones*) is approximately (Sethares, 2005)

$$\ell = (1/16)2^{SPL/10}; \; SPL = 20log_{10}(P_e/P_{ref}); \; P_e = a/\sqrt{2}\,; \; P_{ref} = 20\mu Pa^2, \quad (4)$$

where $SPL$ is the sound pressure level and $P_{ref}$ is the standard reference for the $SPL$ in the air. This model does not include effects such as the Fletcher-Munson curves (Sethares, 2005), and the relation between loudness and amplitude is independent of the frequency: $\ell \cong c_o a^{0.60}$ ($c_o$ constant) (Lawrence & Florentine, 2010). With this approximation $\ell_{min}$ is found to be proportional to the minimum value between ${a_1}^{0.60}$ and ${a_2}^{0.60}$. In a normalized scale, the loudness can be measured using the normalized amplitude $a_{norm}$, leading to $\ell_{norm} = \ell/\ell_{max} = (a/a_{max})^{0.60} = (a_{norm})^{0.60}$.

For a complex tone with $m$ partials, the total dissonance $D_F$ can be calculated by superposing the individual dissonances for each possible pair of partials, it is

$$D_F = \frac{1}{2}\sum_{i=1}^{m}\sum_{j=1}^{m} d(f_i, f_j, x_i, x_j), \quad (5)$$

with $x_k = a_k$ for the first model of Sethares (1998) and for the Vassilakis model, and $x_k = \ell_k$ for the second model of Sethares (2005).

**Dissonance level of pairs of complex tones with the same timbre.** If we superpose two different complex tones with the same timbre and with a ratio between their fundamental frequencies of $f_k/f_l = \alpha$, with $f_k > f_l$, then the total dissonance of this superposition $D_F(\alpha)$ is given by (Sethares, 1993, 1998)

$$D_F(\alpha) = D_F + D_{\alpha F} + \sum_{i=1}^{m} \sum_{j=1}^{m} d(f_i, \alpha f_j, x_i, x_j), \quad (6)$$

where $D_F$ and $D_{\alpha F}$ correspond to the total dissonance associated with the timbre of each individual complex tone, and the last term is due to the total dissonance generated by the interaction between the partials of the two complex tones. Here the amplitudes and the relative distances between the partials in the spectrum are the same, independent of the fundamental frequency, and this equation can be used to identify the local minima of dissonance by scanning all possible values of $\alpha$ for the pair of fundamental frequencies $f_k$ and $f_l$. Usually the lowest fundamental frequency is taken as fixed, and the highest one varies—changing the values of $\alpha$.

In the case of equal amplitudes and for a normalized scale of dissonance, the first Sethares model and the Vassilakis model give the same results; additionally, if the Stevens law is used in the second Sethares model, then the three models are equivalent.

**Determination of the fundamental frequency for the Miñana study.** In order to compare the results from this study with those from a previous one carried out in 1990 by Carlos Miñana (2010a), including 9 traditional marimbas made between 1950 and 1986, the fundamental frequency of each bar was obtained for all the marimbas.

Miñana produced a score with the approximate pitches associated with each marimba bar, including their corresponding increment or decrement in cents with respect to the TTET scale with $A = 440\ Hz$. The approximate fundamental frequencies ($f_{Real}$) produced by these marimbas were found from the fundamental frequencies in the TTET scale ($f_{Temp}$), and using the relation between the number of cents $e$ and the corresponding frequency ratio $r$: $f_{Real} \approx r \cdot f_{Temp}$; $r = c^e$, where $c$ refers to a cent ($c = 2^{1/1200}$) (Roederer, 2009). Supplement 1 contains the score as presented by Miñana (2010a).

## Experimental

In order to obtain evidence about the musical practices related to the traditional marimba, as well as about the tunings produced by instrument makers, an expedition to the heart of the *marimba de chonta* territories was carried out in 2015. The expedition was composed by musicians and researchers from the Conservatory of Music, the Department of Physics, and the School of Cinema and Media Arts of Universidad Nacional de Colombia. During the expedition the researchers interviewed instrument makers, female singers, and *marimba de chonta* interpreters, as well as music teachers from local schools. Additionally, the team attended several presentations by local musical groups.

**Selection of the marimbas.** This study includes ten traditional marimbas, each one constructed by a different recognized instrument maker, and one diatonic *marimba de chonta* tuned in the TTET. Table 1 shows the instrument makers and the main features of the selected marimbas. Marimbas 1, 2, 3, 4, 6, 9, and 11 were made between 2014 and 2015. The remaining ones are undated and presumably old instruments; however, these marimbas are currently in use, and a common practice is to repair these instruments due to their natural fragility and the harsh conditions of use.

| Marimba number | Maker | Place | Bars | Kind |
|---|---|---|---|---|
| 1 | Baudilio Cuama | Buenaventura | 22 | Tempered |
| 2 | Baudilio Cuama | Buenaventura | 16 | Traditional |
| 3 | Jhon Jairo Cortés | Tumaco | 24 | Traditional |
| 4 | Francisco Tenorio | Tumaco | 24 | Traditional |
| 5 | Juan E. Sinisterra | Tumaco | 18 | Traditional |
| 6 | Silvino Mina | Guapi | 20 | Traditional |
| 7 | Dioselino Rodríguez | Guapi | 17 | Traditional |
| 8 | Guillermo Ríos | Guapi | 17 | Traditional |
| 9 | Genaro Torres | Guapi | 21 | Traditional |
| 10 | José Torres | Guapi | 20 | Traditional |
| 11 | Francisco Torres | Guapi | 24 | Traditional |

**Table 1.** Description of each marimba recorded for this study, including identification number, instrument maker, place, total number of bars, and tuning.

**Procedure for the analysis of pitch and timbre.** For all marimbas the sound of each bar with its respective resonator were recorded; in all cases the bars were struck in the geometrical center in agreement with the common style of traditional musicians. The spectrum was obtained for samples covering the sound between the attack and the final release. The Fast Fourier Transform over a Hanning window was used for finding the frequencies and amplitudes of the spectrum of each bar. Then, the peaks corresponding to the fundamental and the first ten overtones with the largest amplitudes were identified.

For almost every bar, the fundamental frequency corresponds to the peak with the largest amplitude; however, for the traditional marimbas of the Torres family (numbers 9, 10, and 11), some bars have overtones with larger amplitudes than that of the fundamental. In the sample of 223 bars, the fundamental frequency was not clearly identified for only four bars (marked with "*" in Supplement 2); these values were not used in the analysis. For all bars of each marimba the amplitudes were normalized for the analysis.

**Procedure for the analysis of musical pieces.** Seven musical pieces played by recognized traditional marimba musicians were recorded, with one musician and two mallets for each marimba (a maximum of two pitches can be sounded simultaneously in each musical piece). Musical scores with the closest transcriptions to the TTET scale with $A = 440\ Hz$ were generated. Traditional tunings do not follow a TTET system; however, this procedure allows the identification of the approximate size of harmonic intervals in order to measure their frequency of occurrence. The musical scores are provided in Supplement 3.

In order to infer the most frequent harmonic intervals in traditional music, the musical scores were used to measure the probability of occurrence of each interval size. In order to distinguish between brief and lengthy intervals, a second analysis was carried out taking into account the time duration of each kind of interval, according to its size. The probability of occurrence was defined as proportional to the sum of the duration of each interval of a given size. For $u$ different intervals with size $z$ and $F_z$ occurrences for each size the probability of a specific interval size is $p_z = T_z/T$, where $T_z = t_{1_z} + t_{2_z} + \cdots + t_{F_z}$ is the sum of the durations of all intervals with size $z$ and total time $T = T_1 + T_2 + \cdots + T_u$. If all intervals have the same duration in a musical score, then the probability $p_z$ is equal to the probability of occurrence.

## Results and Discussion

### Tunings

The fundamental frequency of each marimba bar coupled to its respective resonator was obtained (Supplement 2). The TTET marimba was found to be tuned in a major diatonic tempered scale over $C$ with the traditional reference $A = 440\ Hz$ (Supplement 2).

For the traditional marimbas, the average frequency ratios that result from the combination of all bars at different distances, the minimum and maximum values, and the Standard Deviation found for each marimba in the study carried out by Miñana (2010a) and in the present study are presented in Tables 2 and 3, respectively. The distance $s$ between bars is defined as the number of steps to reach the final bar starting from the initial one; hence, adjacent bars have a distance of 1 step.

The traditional marimbas studied by Miñana are tuned using equi-heptatonic scales; he reports this tendency, and this can also be inferred from the new analysis carried out in the present research. For each marimba, the average ratio of the fundamental frequencies for bars separated by a distance of 7 steps ($r_7$) is slightly smaller than two, $1.91 \leq r_7 \leq 1.98$, with an average of 1.94 (see Table 2); hence, these scales are defined by low octaves, in agreement with the findings of Miñana. The equi-heptatonic scale fulfills the condition that the frequency ratio between two bars with a distance $s$ is ${r_7}^{s/7}$. The difference between these equi-heptatonic scales is the value of $r_7$. The theoretical equi-heptatonic scales constructed using the experimental values of $r_7$ are presented in Table 4.

| Distance (steps) | | Marimba number | | | | | | | | | Avg.* |
|---|---|---|---|---|---|---|---|---|---|---|---|
| | | 1 | 2 | 3 | 4 | 5 | 6 | 7 | 8 | 9 | |
| 1 | **Min** | 1.04 | 1.06 | 1.08 | 1.08 | 1.07 | 1.04 | 1.07 | 1.06 | 1.05 | 1.06 |
| | **Max** | 1.16 | 1.16 | 1.14 | 1.12 | 1.13 | 1.12 | 1.15 | 1.15 | 1.17 | 1.14 |
| | **Avg.** | 1.11 | 1.10 | 1.10 | 1.10 | 1.10 | 1.09 | 1.10 | 1.10 | 1.09 | 1.10 |
| | **σ** | 0.03 | 0.03 | 0.01 | 0.01 | 0.01 | 0.03 | 0.02 | 0.02 | 0.03 | 0.02 |
| 2 | **Min** | 1.18 | 1.19 | 1.18 | 1.19 | 1.14 | 1.11 | 1.16 | 1.15 | 1.11 | 1.16 |
| | **Max** | 1.30 | 1.27 | 1.23 | 1.23 | 1.24 | 1.25 | 1.25 | 1.26 | 1.26 | 1.26 |
| | **Avg.** | 1.22 | 1.22 | 1.21 | 1.21 | 1.20 | 1.19 | 1.21 | 1.20 | 1.20 | 1.21 |
| | **σ** | 0.03 | 0.03 | 0.02 | 0.01 | 0.03 | 0.04 | 0.02 | 0.03 | 0.03 | 0.03 |
| 3 | **Min** | 1.32 | 1.27 | 1.29 | 1.30 | 1.24 | 1.18 | 1.29 | 1.26 | 1.22 | 1.26 |
| | **Max** | 1.42 | 1.41 | 1.35 | 1.36 | 1.36 | 1.40 | 1.38 | 1.37 | 1.41 | 1.38 |
| | **Avg.** | 1.35 | 1.34 | 1.33 | 1.33 | 1.32 | 1.31 | 1.34 | 1.32 | 1.31 | 1.33 |
| | **σ** | 0.03 | 0.04 | 0.02 | 0.01 | 0.03 | 0.05 | 0.03 | 0.03 | 0.05 | 0.03 |
| 4 | **Min** | 1.41 | 1.46 | 1.42 | 1.43 | 1.36 | 1.31 | 1.41 | 1.37 | 1.32 | 1.39 |
| | **Max** | 1.60 | 1.52 | 1.51 | 1.50 | 1.48 | 1.52 | 1.52 | 1.53 | 1.56 | 1.53 |
| | **Avg.** | 1.48 | 1.48 | 1.46 | 1.46 | 1.46 | 1.44 | 1.47 | 1.46 | 1.44 | 1.46 |
| | **σ** | 0.05 | 0.02 | 0.02 | 0.02 | 0.03 | 0.05 | 0.03 | 0.04 | 0.06 | 0.04 |
| 5 | **Min** | 1.57 | 1.57 | 1.56 | 1.58 | 1.53 | 1.47 | 1.55 | 1.53 | 1.46 | 1.54 |
| | **Max** | 1.71 | 1.68 | 1.64 | 1.64 | 1.63 | 1.66 | 1.68 | 1.70 | 1.68 | 1.67 |
| | **Avg.** | 1.63 | 1.63 | 1.60 | 1.61 | 1.60 | 1.58 | 1.62 | 1.60 | 1.59 | 1.61 |
| | **σ** | 0.04 | 0.04 | 0.03 | 0.02 | 0.03 | 0.05 | 0.04 | 0.05 | 0.07 | 0.04 |
| 6 | **Min** | 1.74 | 1.75 | 1.72 | 1.73 | 1.68 | 1.64 | 1.72 | 1.66 | 1.61 | 1.69 |
| | **Max** | 1.91 | 1.87 | 1.79 | 1.79 | 1.80 | 1.84 | 1.86 | 1.84 | 1.86 | 1.84 |
| | **Avg.** | 1.80 | 1.80 | 1.76 | 1.76 | 1.76 | 1.74 | 1.78 | 1.75 | 1.74 | 1.77 |
| | **σ** | 0.05 | 0.04 | 0.02 | 0.02 | 0.03 | 0.06 | 0.04 | 0.05 | 0.08 | 0.04 |
| 7 | **Min** | 1.89 | 1.91 | 1.90 | 1.90 | 1.83 | 1.78 | 1.87 | 1.86 | 1.75 | 1.85 |
| | **Max** | 2.13 | 2.05 | 1.97 | 1.97 | 1.99 | 2.01 | 2.04 | 1.99 | 2.05 | 2.02 |
| | **Avg.** | 1.98 | 1.98 | 1.93 | 1.94 | 1.93 | 1.91 | 1.97 | 1.92 | 1.92 | 1.94 |
| | **σ** | 0.07 | 0.05 | 0.02 | 0.02 | 0.04 | 0.07 | 0.05 | 0.04 | 0.09 | 0.05 |

**Table 2.** Minimum, maximum, average, and Standard Deviation of the frequency ratio for pairs of bars separated by different distances in the traditional marimbas recorded by Miñana (2010). The numeration for the marimbas presented by Miñana (2010) is conserved. The symbol "σ" refers to the Standard Deviation. Avg. refers to the average. The average with the symbol "*" refers to the average of the minimum, the maximum, the average, and σ values for all marimbas.

The traditional marimbas recorded for the present study have three different behaviors (Table 3): 7 marimbas numbered from 2 up to 8 were found to be equi-heptatonic, with $1.96 \leq r_7 \leq 2.03$ and average of 2.00; marimbas 9 and 10 follow an equi-octatonic scale ${r_8}^{s/8}$ ($r_8 = 2.04$ and $r_8 = 2.08$); and the marimba number 11 follows an equi-enneatonic scale with $r_9 = 1.99$. For this case, $r_7$, $r_8$ and $r_9$ refer to the average ratios of the fundamental frequencies for bars separated by a distance of 7, 8 and 9 steps, respectively. The theoretical equi-heptatonic, equi-octatonic and equi-enneatonic scales constructed using $r_7$, $r_8$ and $r_9$ are presented in Table 5.

For all the traditional marimbas studied the predicted values presented in Tables 4, 5 have less than 1.00 % relative error with respect to the empirical values shown in Tables 2, 3.

| Distance (steps) | | Marimba number | | | | | | | | | | | |
|---|---|---|---|---|---|---|---|---|---|---|---|---|---|
| | | 2 | 3 | 4 | 5 | 6 | 7 | 8 | Avg.*$_1$ | 9 | 10 | Avg.*$_2$ | 11 |
| 1 | **Min** | 1.08 | 1.06 | 1.05 | 1.07 | 1.02 | 1.07 | 1.06 | 1.06 | 1.05 | 0.99 | 1.02 | 1.04 |
| | **Max** | 1.15 | 1.17 | 1.15 | 1.14 | 1.15 | 1.14 | 1.15 | 1.15 | 1.16 | 1.15 | 1.16 | 1.15 |
| | **Avg.** | 1.11 | 1.10 | 1.10 | 1.11 | 1.10 | 1.10 | 1.11 | 1.10 | 1.09 | 1.09 | 1.09 | 1.08 |
| | **σ** | 0.02 | 0.03 | 0.02 | 0.02 | 0.03 | 0.02 | 0.02 | 0.02 | 0.03 | 0.04 | 0.04 | 0.03 |
| 2 | **Min** | 1.19 | 1.16 | 1.15 | 1.19 | 1.16 | 1.19 | 1.15 | 1.17 | 1.15 | 1.15 | 1.15 | 1.09 |
| | **Max** | 1.28 | 1.27 | 1.29 | 1.28 | 1.26 | 1.27 | 1.26 | 1.27 | 1.24 | 1.27 | 1.26 | 1.23 |
| | **Avg.** | 1.22 | 1.22 | 1.22 | 1.23 | 1.21 | 1.21 | 1.22 | 1.22 | 1.20 | 1.20 | 1.20 | 1.17 |
| | **σ** | 0.03 | 0.03 | 0.04 | 0.02 | 0.03 | 0.02 | 0.03 | 0.03 | 0.02 | 0.03 | 0.03 | 0.04 |
| 3 | **Min** | 1.31 | 1.28 | 1.28 | 1.31 | 1.29 | 1.28 | 1.24 | 1.28 | 1.25 | 1.20 | 1.23 | 1.17 |
| | **Max** | 1.39 | 1.40 | 1.40 | 1.38 | 1.43 | 1.39 | 1.41 | 1.40 | 1.40 | 1.41 | 1.41 | 1.32 |
| | **Avg.** | 1.35 | 1.35 | 1.34 | 1.35 | 1.33 | 1.34 | 1.34 | 1.34 | 1.31 | 1.31 | 1.31 | 1.26 |
| | **σ** | 0.02 | 0.03 | 0.03 | 0.02 | 0.03 | 0.03 | 0.05 | 0.03 | 0.04 | 0.05 | 0.05 | 0.05 |
| 4 | **Min** | 1.45 | 1.41 | 1.42 | 1.45 | 1.42 | 1.42 | 1.37 | 1.42 | 1.38 | 1.38 | 1.38 | 1.26 |
| | **Max** | 1.53 | 1.54 | 1.53 | 1.55 | 1.52 | 1.52 | 1.55 | 1.53 | 1.48 | 1.53 | 1.51 | 1.41 |
| | **Avg.** | 1.49 | 1.48 | 1.49 | 1.50 | 1.47 | 1.47 | 1.48 | 1.48 | 1.43 | 1.44 | 1.44 | 1.36 |
| | **σ** | 0.03 | 0.03 | 0.04 | 0.03 | 0.03 | 0.03 | 0.06 | 0.03 | 0.04 | 0.04 | 0.04 | 0.03 |
| 5 | **Min** | 1.57 | 1.56 | 1.56 | 1.61 | 1.55 | 1.57 | 1.51 | 1.56 | 1.50 | 1.48 | 1.49 | 1.40 |
| | **Max** | 1.70 | 1.73 | 1.76 | 1.72 | 1.69 | 1.68 | 1.72 | 1.71 | 1.66 | 1.70 | 1.68 | 1.54 |
| | **Avg.** | 1.64 | 1.64 | 1.65 | 1.66 | 1.62 | 1.62 | 1.64 | 1.64 | 1.56 | 1.58 | 1.57 | 1.46 |
| | **σ** | 0.04 | 0.04 | 0.05 | 0.04 | 0.05 | 0.04 | 0.06 | 0.04 | 0.04 | 0.06 | 0.05 | 0.04 |
| 6 | **Min** | 1.76 | 1.73 | 1.74 | 1.78 | 1.70 | 1.73 | 1.71 | 1.74 | 1.64 | 1.64 | 1.64 | 1.51 |
| | **Max** | 1.87 | 1.89 | 1.93 | 1.87 | 1.91 | 1.90 | 1.89 | 1.89 | 1.77 | 1.82 | 1.80 | 1.69 |
| | **Avg.** | 1.82 | 1.81 | 1.82 | 1.84 | 1.78 | 1.79 | 1.81 | 1.81 | 1.71 | 1.73 | 1.72 | 1.59 |
| | **σ** | 0.04 | 0.04 | 0.05 | 0.02 | 0.05 | 0.04 | 0.06 | 0.04 | 0.05 | 0.05 | 0.05 | 0.06 |
| 7 | **Min** | 1.98 | 1.92 | 1.95 | 1.98 | 1.92 | 1.88 | 1.89 | 1.93 | 1.80 | 1.77 | 1.79 | 1.58 |
| | **Max** | 2.04 | 2.08 | 2.10 | 2.07 | 2.05 | 2.03 | 2.08 | 2.06 | 1.96 | 2.02 | 1.99 | 1.78 |
| | **Avg.** | 2.01 | 2.00 | 2.00 | 2.03 | 1.96 | 1.97 | 2.01 | 2.00 | 1.87 | 1.90 | 1.89 | 1.71 |
| | **σ** | 0.02 | 0.04 | 0.05 | 0.03 | 0.04 | 0.05 | 0.06 | 0.04 | 0.05 | 0.08 | 0.06 | 0.06 |
| 8 | **Min** | --- | --- | --- | --- | --- | --- | --- | --- | 1.98 | 1.98 | 1.98 | 1.70 |
| | **Max** | --- | --- | --- | --- | --- | --- | --- | --- | 2.14 | 2.17 | 2.16 | 1.95 |
| | **Avg.** | --- | --- | --- | --- | --- | --- | --- | --- | 2.04 | 2.08 | 2.06 | 1.85 |
| | **σ** | --- | --- | --- | --- | --- | --- | --- | --- | 0.06 | 0.07 | 0.06 | 0.06 |
| 9 | **Min** | --- | --- | --- | --- | --- | --- | --- | --- | --- | --- | --- | 1.91 |
| | **Max** | --- | --- | --- | --- | --- | --- | --- | --- | --- | --- | --- | 2.10 |
| | **Avg.** | --- | --- | --- | --- | --- | --- | --- | --- | --- | --- | --- | 1.99 |
| | **σ** | --- | --- | --- | --- | --- | --- | --- | --- | --- | --- | --- | 0.06 |

**Table 3.** Minimum, maximum, average, and Standard Deviation of the frequency ratio for pairs of bars separated by different distances in the traditional marimbas recorded for this study. The symbol "σ" refers to the Standard Deviation. "Avg." refers to the average. "Avg.*$_1$" refers to the average of the minimum, the maximum, the average, and σ values for marimbas 2, 3, 4, 5, 6, 7, and 8. "Avg.*$_2$" refers to the average of the minimum, the maximum, the average, and σ values for marimbas 9, and 10.

| Distance | Marimba number | | | | | | | | | |
|---|---|---|---|---|---|---|---|---|---|---|
| (steps) | 1 | 2 | 3 | 4 | 5 | 6 | 7 | 8 | 9 | Avg.* |
| 1 | 1.10 | 1.10 | 1.10 | 1.10 | 1.10 | 1.10 | 1.10 | 1.10 | 1.10 | 1.10 |
| 2 | 1.22 | 1.22 | 1.21 | 1.21 | 1.21 | 1.20 | 1.21 | 1.20 | 1.20 | 1.21 |
| 3 | 1.34 | 1.34 | 1.33 | 1.33 | 1.33 | 1.32 | 1.34 | 1.32 | 1.32 | 1.33 |
| 4 | 1.48 | 1.48 | 1.46 | 1.46 | 1.46 | 1.45 | 1.47 | 1.45 | 1.45 | 1.46 |
| 5 | 1.63 | 1.63 | 1.60 | 1.61 | 1.60 | 1.59 | 1.62 | 1.59 | 1.59 | 1.61 |
| 6 | 1.80 | 1.80 | 1.76 | 1.76 | 1.76 | 1.74 | 1.79 | 1.75 | 1.75 | 1.76 |
| 7 | 1.98 | 1.98 | 1.93 | 1.94 | 1.93 | 1.91 | 1.97 | 1.92 | 1.92 | 1.94 |

**Table 4.** Theoretical equi-heptatonic scales for the traditional marimbas recorded by Miñana (2010). "Avg.*" refers to the theoretical scale constructed for the average of the average value for all traditional marimbas. In all cases, the values predicted by the theoretical scales have a relative error $\leq$ 1.00 % with respect to the experimental values presented in Table 2.

| Distance | Marimba number (present study) | | | | | | | | | | |
|---|---|---|---|---|---|---|---|---|---|---|---|
| (steps) | 2 | 3 | 4 | 5 | 6 | 7 | 8 | Avg.*$_1$ | 9 | 10 | Avg.*$_2$ | 11 |
| 1 | 1.10 | 1.10 | 1.10 | 1.11 | 1.10 | 1.10 | 1.10 | 1.10 | 1.09 | 1.10 | 1.09 | 1.08 |
| 2 | 1.22 | 1.22 | 1.22 | 1.22 | 1.21 | 1.21 | 1.22 | 1.22 | 1.20 | 1.20 | 1.20 | 1.17 |
| 3 | 1.35 | 1.35 | 1.35 | 1.35 | 1.33 | 1.34 | 1.35 | 1.35 | 1.31 | 1.32 | 1.31 | 1.26 |
| 4 | 1.49 | 1.49 | 1.49 | 1.50 | 1.47 | 1.47 | 1.49 | 1.49 | 1.43 | 1.44 | 1.44 | 1.36 |
| 5 | 1.65 | 1.64 | 1.64 | 1.66 | 1.62 | 1.62 | 1.65 | 1.64 | 1.56 | 1.58 | 1.57 | 1.47 |
| 6 | 1.82 | 1.81 | 1.81 | 1.83 | 1.78 | 1.79 | 1.82 | 1.81 | 1.71 | 1.73 | 1.72 | 1.58 |
| 7 | 2.01 | 2.00 | 2.00 | 2.03 | 1.96 | 1.97 | 2.01 | 2.00 | 1.87 | 1.90 | 1.88 | 1.71 |
| 8 | --- | --- | --- | --- | --- | --- | --- | --- | 2.04 | 2.08 | 2.06 | 1.84 |
| 9 | --- | --- | --- | --- | --- | --- | --- | --- | --- | --- | --- | 1.99 |

**Table 5.** Theoretical equi-heptatonic, equi-octatonic, and equi-enneatonic scales for the traditional marimbas recorded for this study. "Avg.*$_1$" refers to the theoretical scale constructed for the average of the average values over the marimbas 2,3,4,5,6,7 and 8. "Avg.*$_2$" refers to the theoretical scale constructed for the average of the averages values of the marimbas 9 and 10. In all cases, the values predicted by the theoretical scales have a relative error $\leq$ 1.00 % with respect to the experimental values presented in Table 3.

The three traditional marimbas with equi-octatonic and equi-enneatonic scales were made by three members of the Torres family, two of these makers also made three of the marimbas studied by Miñana (marimbas 1, 2, 5 in Table 2) that were found to be equi-heptatonic. In all cases, the standard deviation of the frequency ratios for all the possible combinations of each distance between bars is smaller than 5.00 %, with respect to the average value (Tables 2, 3). This is presumably due to variations in the voice preferences of female singers in a region or territory. Since in each territory there are several singers that are accompanied by musicians playing the same traditional marimba, it is impossible to define only one fundamental frequency as a reference to construct the scale. Hence, the marimbas are constructed by adjusting the tuning as close as possible to the different vocal preferences (Duque, Sanchez & Tascón, 2009), but approximately preserving an isotonic tuning in order to keep the geometrical distances in a transposition process.

The results of this section show that traditional tunings have changed with time in the last three decades, losing the low octaves feature in favor of octaves usually closer to the just interval. This change may be the result of the influence of Western music, which is characterized by the just octave. It has been argued that the globalization of Western music, and the commercial value of new genres that combine local rhythms and timbres, have led to a broad use of TTET *marimbas de chonta*, modifying the taste of African descendants in the Pacific Coast of Colombia (Hernández, 2007). Frequently the traditional marimbas are referred to as "badly made marimbas" or "wrongly tuned marimbas". The most important cultural event of music from this region, the *Petronio Alvarez Festival* (Hernández, 2007), promoted the use of TTET marimbas for several years. It was not until 2015 that the festival opened a category for music played with traditional marimbas. It is a common practice of the Ministry of Culture of Colombia to supply the music schools with TTET marimbas, where music is taught using the logics of diatonic and chromatic scales with tones and semitones.

Finally, these results also show the use of equi-octatonic and equi-enneatonic scales, and that the equal distances feature of the scales has been preserved.

**Use of harmonic intervals in the musical practices**

According to Miñana (2010a), neutral thirds are among the most used harmonic intervals in this traditional music. Frequency ratios covering the region of the thirds are commonly generated using distances of 2 and 3 steps between bars (see Tables 2, 3). In order to check the frequency of use of intervals, their probability of occurrence was obtained for seven performances of traditional marimba pieces by applying the procedure described in the Methods section. If one of the most used intervals is the neutral third, then minor and major thirds are expected to occur with high probability in the scores.

Table 6 shows the probability of occurrence of each harmonic interval, and the probability of occurrence proportional to the total time of duration. Intervals between 3 and 4 semitones are present with high probability in all pieces. Intervals containing sizes between the fourths and fifths, in most cases generated by pairs of bars separated by a distance of 3 and 4 steps, are also frequent in some pieces; and those intervals near to the sixths, in most cases generated by 5 steps, appear less frequently in performances. The region of the seconds, generated by a distance of 1 step, is completely avoided in all musical pieces, and intervals larger than one octave are almost absent. The region of the sevenths, in most cases generated by a distance of 6 steps, is avoided in all pieces with the exception of one, where it has a small incidence, and the region of the octaves is used with small probability of occurrence.

The geometrical schemes shown in Figure 1 can be applied for the traditional marimbas with equi-heptatonic scales, the most common scales found in the present study, and in the studies of Miñana. From these schemes it is possible to infer that the harmonic intervals with distances of 1 and 6 steps between bars are the least frequent. Figure 2 shows the relation between the total number of different combinations of bars and the distance between them for marimbas with different sizes, for the cases of the hexatonic and the pentatonic scales. Supplement 4 contains the exact number of combinations for each scale and marimba size. Notice that adjacent bars and pairs of bars separated by 6 steps are highly uncommon, this is due to the harmony rules of the geometrical schemes and incidence of the discarded bars, the so called “bad bars”.

| Interval (semitones) | Probability | | | | | | | | | | | | | | | |
|---|---|---|---|---|---|---|---|---|---|---|---|---|---|---|---|---|
| | Frequency of occurrence | | | | | | | | Total duration | | | | | | | |
| | Musical piece | | | | | | | | Musical piece | | | | | | | |
| | 1 | 2 | 3 | 4 | 5 | 6 | 7 | Avg. | 1 | 2 | 3 | 4 | 5 | 6 | 7 | Avg. |
| **1** | 0.00 | 0.00 | 0.00 | 0.00 | 0.00 | 0.00 | 0.00 | 0.00 | 0.00 | 0.00 | 0.00 | 0.00 | 0.00 | 0.00 | 0.00 | 0.00 |
| **2** | 0.00 | 0.00 | 0.00 | 0.00 | 0.00 | 0.00 | 0.00 | 0.00 | 0.00 | 0.00 | 0.00 | 0.00 | 0.00 | 0.00 | 0.00 | 0.00 |
| **3** | 0.24 | 0.59 | 0.53 | 0.13 | 0.55 | 0.23 | 0.55 | 0.40 | 0.19 | 0.60 | 0.52 | 0.13 | 0.56 | 0.23 | 0.54 | 0.39 |
| **4** | 0.08 | 0.41 | 0.47 | 0.12 | 0.44 | 0.26 | 0.37 | 0.31 | 0.08 | 0.40 | 0.48 | 0.12 | 0.43 | 0.26 | 0.35 | 0.30 |
| **5** | 0.23 | 0.00 | 0.00 | 0.30 | 0.00 | 0.34 | 0.07 | 0.13 | 0.26 | 0.00 | 0.00 | 0.33 | 0.00 | 0.34 | 0.07 | 0.14 |
| **6** | 0.00 | 0.00 | 0.00 | 0.01 | 0.00 | 0.00 | 0.02 | 0.00 | 0.00 | 0.00 | 0.00 | 0.01 | 0.00 | 0.00 | 0.03 | 0.01 |
| **7** | 0.07 | 0.00 | 0.00 | 0.17 | 0.01 | 0.15 | 0.00 | 0.06 | 0.05 | 0.00 | 0.00 | 0.20 | 0.00 | 0.15 | 0.00 | 0.06 |
| **8** | 0.21 | 0.00 | 0.00 | 0.03 | 0.00 | 0.00 | 0.00 | 0.03 | 0.17 | 0.00 | 0.00 | 0.02 | 0.00 | 0.00 | 0.00 | 0.03 |
| **9** | 0.13 | 0.00 | 0.00 | 0.02 | 0.00 | 0.00 | 0.00 | 0.02 | 0.20 | 0.00 | 0.00 | 0.02 | 0.00 | 0.00 | 0.00 | 0.03 |
| **10** | 0.00 | 0.00 | 0.00 | 0.15 | 0.00 | 0.00 | 0.00 | 0.02 | 0.01 | 0.00 | 0.00 | 0.12 | 0.00 | 0.00 | 0.00 | 0.02 |
| **11** | 0.00 | 0.00 | 0.00 | 0.00 | 0.00 | 0.00 | 0.00 | 0.00 | 0.00 | 0.00 | 0.00 | 0.00 | 0.00 | 0.00 | 0.00 | 0.00 |
| **12** | 0.05 | 0.00 | 0.00 | 0.07 | 0.00 | 0.02 | 0.00 | 0.02 | 0.04 | 0.00 | 0.00 | 0.06 | 0.00 | 0.02 | 0.00 | 0.02 |
| **>12** | 0.00 | 0.00 | 0.00 | 0.01 | 0.00 | 0.00 | 0.00 | 0.00 | 0.00 | 0.00 | 0.00 | 0.01 | 0.00 | 0.00 | 0.00 | 0.00 |

**Table 6.** Probability of occurrence of harmonic intervals as a function of their size in semitones. Avg. refers to the average of the seven musical pieces. The probability has been taken to be proportional to the frequency of occurrence and to the total time of duration of each size of harmonic interval.

Results of the analysis of the frequency ratios of the traditional marimbas with equi-heptatonic scales are presented in Tables 2 and 3. These regions correspond to intervals around the seconds (frequency ratios between or near to the range $[1.06, 1.12]$) and the sevenths (frequency ratios between or near to the range $[1.78, 1.89]$).

For the same distance between bars, the marimbas of the Torres family analyzed in the present study have lower values in the tunings in comparison with the other ones (see Table 3), this feature is more evident for larger distances than for smaller ones (a natural consequence of the presence of equi-octatonic and equi-enneatonic scales). For marimbas 9 and 10 (see Table 3), intervals near to the sevenths and the octaves are better generated using 7 and 8 steps, respectively. For marimba 11 (see Table 3), the same intervals are better generated using 8 and 9 steps, respectively.

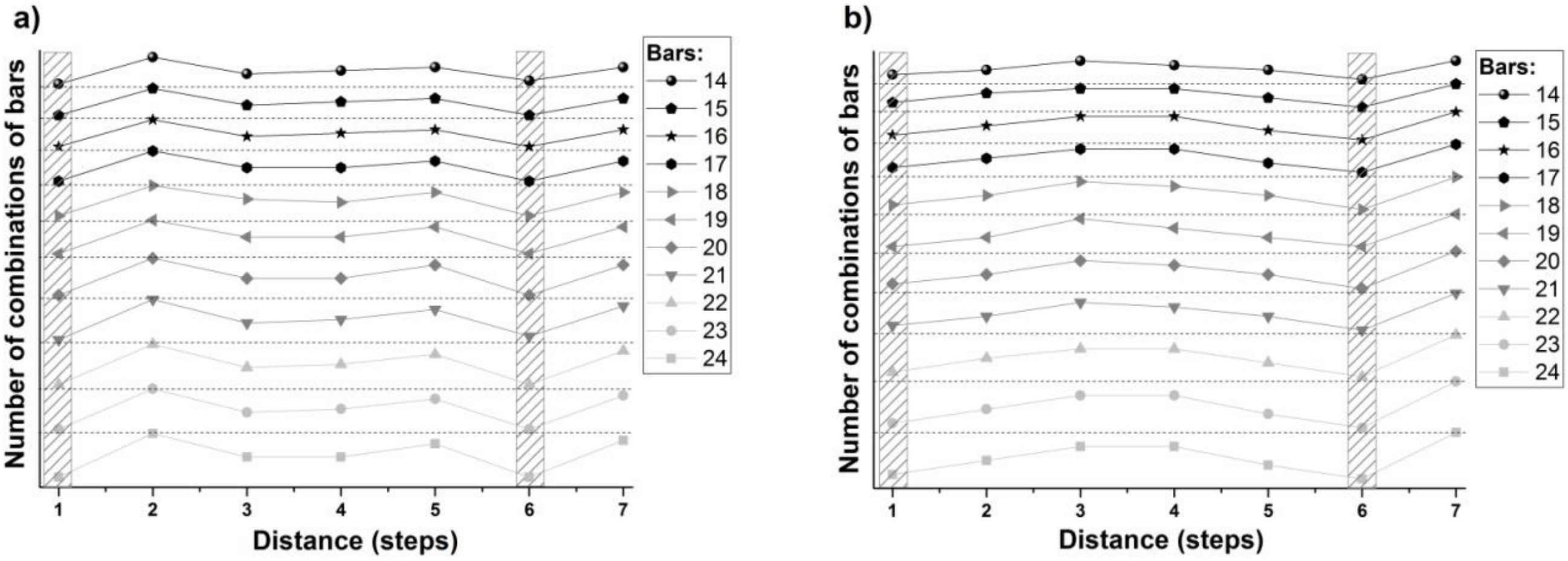

**Figure 2.** Relation between the total number of different combinations of bars and the distance between them, for marimbas of different sizes (from 14 up to 24 bars). The height of each sequence is proportional to the total number of different combinations of pairs of bars. Grey regions show the regions with the lowest number of combinations, which correspond to 1 and 6 steps distances. a) Hexatonic scale. b) Pentatonic scale. In all cases the total number of combinations has been counted beginning with the first bars shown in Figure 1 for the hexatonic and the pentatonic scales.

Finally, from the statistical analysis as well as from the analysis of the geometrical schemes, it is possible to infer that regions around the seconds and the sevenths seem to be deliberately avoided in this music. Since octaves present a small probability of occurrence in the statistical analysis (in average over all the musical pieces, octaves are 2% of all intervals), we infer that the most used region of the traditional marimba music is between the thirds and the sixths (Table 6). In the equi-heptatonic traditional marimbas of this study, and from the study by Miñana, this region is generated using bars separated between 2 and 5 steps; from the average values of the minima and maxima of both studies, this region is approximately between $1.16 \leq \alpha \leq 1.71$ (Tables 2, 3).

## Theoretical dissonance level curves

In this subsection the theoretical dissonance curves associated with the sound emitters of the *marimba de chonta* are presented.

**Isolated bars.** In the tuning process, some makers construct the bars before the resonators and test the tuning within sets of three, four, or five adjacent bars. In order to analyze this procedure, the dissonance curve for an isolated rectangular bar was obtained theoretically. The range used by the instrument makers to test the tuning include fundamental frequency ratios up to $\alpha = 1.53$, which is the maximum value among the average of the maxima found for 5 successive bars (4 steps) in both studies (see Tables 2, 3).

Assuming that the main contributions to the spectrum are from transverse modes of oscillation of a rectangular bar with free ends, the first five overtones are given by $\{2.758f_1, 5.406f_1, 8.936f_1, 13.350f_1, 18.645f_1\}$, where $f_1$ is the fundamental frequency (Fletcher & Rossing, 1998; Kinsler, Frey, Coppens & Sanders, 1999). Figure 3a presents the dissonance curve for this case inside one octave, taking equal values of the amplitudes for the fundamental and the overtones in order to appreciate the contributions of all overtones, in agreement with the procedure carried out by Sethares (1993, 1998, 2005). This curve was reported by Sethares (1993) in order to expose that if the tuning of instruments is carried out according to the consonance properties of their timbre, then bar instruments must be tuned in a different way than instruments with harmonic timbre, such as strings and pipes.

Local minima of dissonance were found at $\alpha = 1.26, 1.40$ and 1.49, as illustrated in Figure 3a. The minimum at $\alpha = 1.40$ does not correspond to any average in the experimental tunings (see Tables 2, 3). The minimum at $\alpha = 1.49$ (close to the just fifth and the TTET fifth) is near the average value for the distance of 4 steps in the case of the equi-heptatonic marimbas in the present study ($\alpha = 1.48$), and slightly farther from the average for the marimbas studied by Miñana ($\alpha = 1.46$). The minimum located at 1.26 is broader than the other minima and covers the region around the thirds and the fourths ($1.16 < \alpha < 1.37$, grey box in Figure 3a). The minimum at 1.26 is less dissonant than the ones found in the same region for the case of a harmonic spectrum (Figure 3a), and it is placed in an important range of use of harmonic intervals in traditional marimba music (see 3, 4, 5 semitones in Table 6).

Figure 3b shows the differentiation of the dissonance level $D_F(\alpha)$ with respect to the $\alpha$ parameter, $dD_F(\alpha)/d\alpha$, illustrating effects due to the shapes of local minima. In the region of the thirds and the fourths the level of dissonance for bars changes smoothly when compared to the curve for emitters with harmonic spectrum.

**Bars and resonators.** In order to include effects due to the tubular resonator located under each bar, it is modeled as a cylindrical closed pipe having the same fundamental frequency as the bar. Since closed pipes only produce odd harmonics (Fletcher & Rossing, 1998), the first thirteen overtones for the bar-resonator system used in the analysis are $\{2.758f_1, 3.000f_1, 5.000f_1, 5.406f_1, 7.000f_1, 8.936f_1, 9.000f_1, 11.000f_1, 13.000f_1, 13.350f_1, 15.000f_1, 17.000f_1, 18.645f_1\}$.

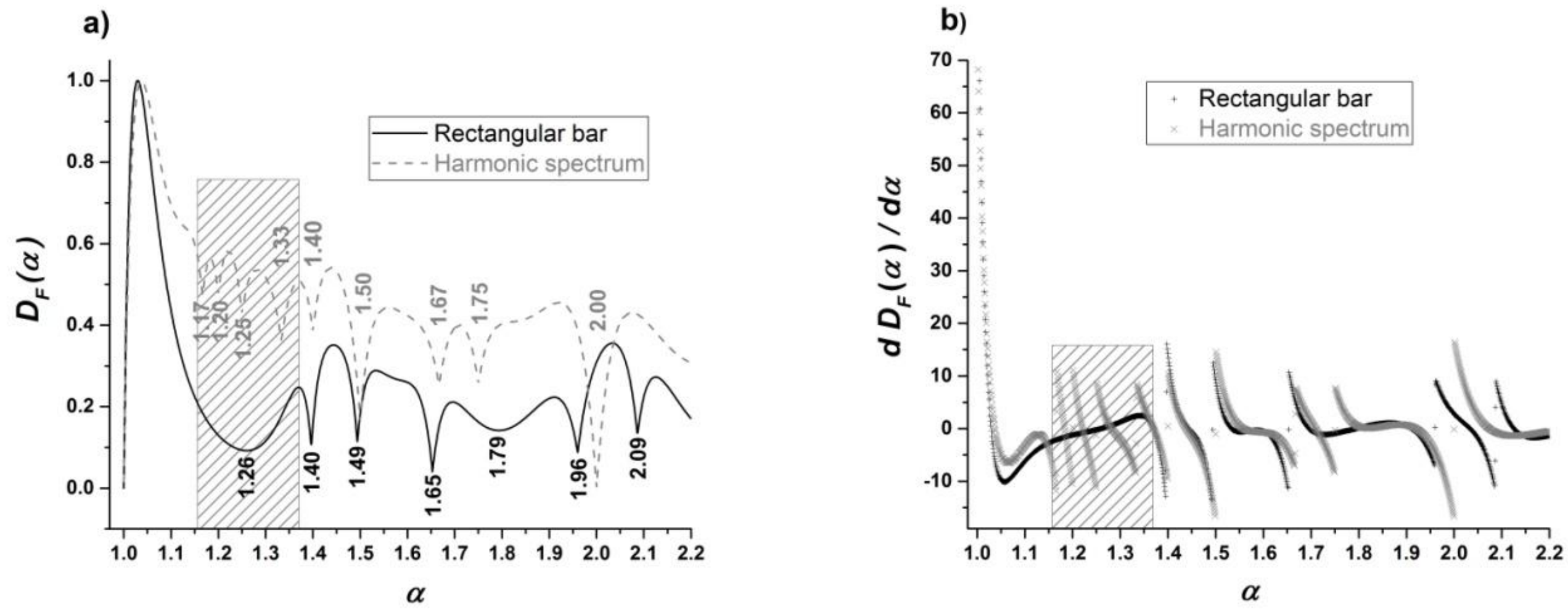

**Figure 3.** a) Dissonance level $D_F(\alpha)$ as a function of the ratio of the fundamental frequencies $\alpha$, for a rectangular bar free to vibrate at both ends, and for the case of a six harmonics spectrum with equal amplitudes. The location of the minima for the harmonic spectrum corresponds to the ratio of small natural numbers posted by Pythagoras: 2/1=2.00, 3/2=1.50, 4/3=1.33, 5/3=1.67, 5/4=1.25, 6/5=1.20, 7/4=1.75, 7/5=1.40, 7/6=1.17. b) Differentiation of the dissonance level $D_F(\alpha)$ with respect to the $\alpha$ parameter, as a function of the ratio between the fundamental frequencies $\alpha$. In both graphics 300 $Hz$ has been taken as the lowest fundamental frequency, and the dissonance level scale has been normalized between 0 and 1.

The amplitudes for the partials are assumed to diminish as $A_n = A_0(0.16)exp[(-1/5)\,(n - 2.758)]$, where $n$ is the ratio of the frequency of the corresponding overtone to the fundamental ($n = 2.758,\ 3.000,\ 5.000, ...$), and $A_0$ is the amplitude for the fundamental. Figure 4 shows the corresponding spectrum.

According to Sethares, the precise value of the amplitude for each overtone can be somewhat arbitrary. However, these values must reflect the average properties of the samples (Sethares, 2005). In this sense, the assumption that the amplitudes $A_n$ diminish exponentially is in agreement with the most common tendency of high overtones having smaller amplitudes than lower ones, and it is in agreement with the average amplitude proportions between the fundamental and the main modes measured experimentally for a set of traditional marimbas (Figure 7). This is discussed in more detail in the following section.

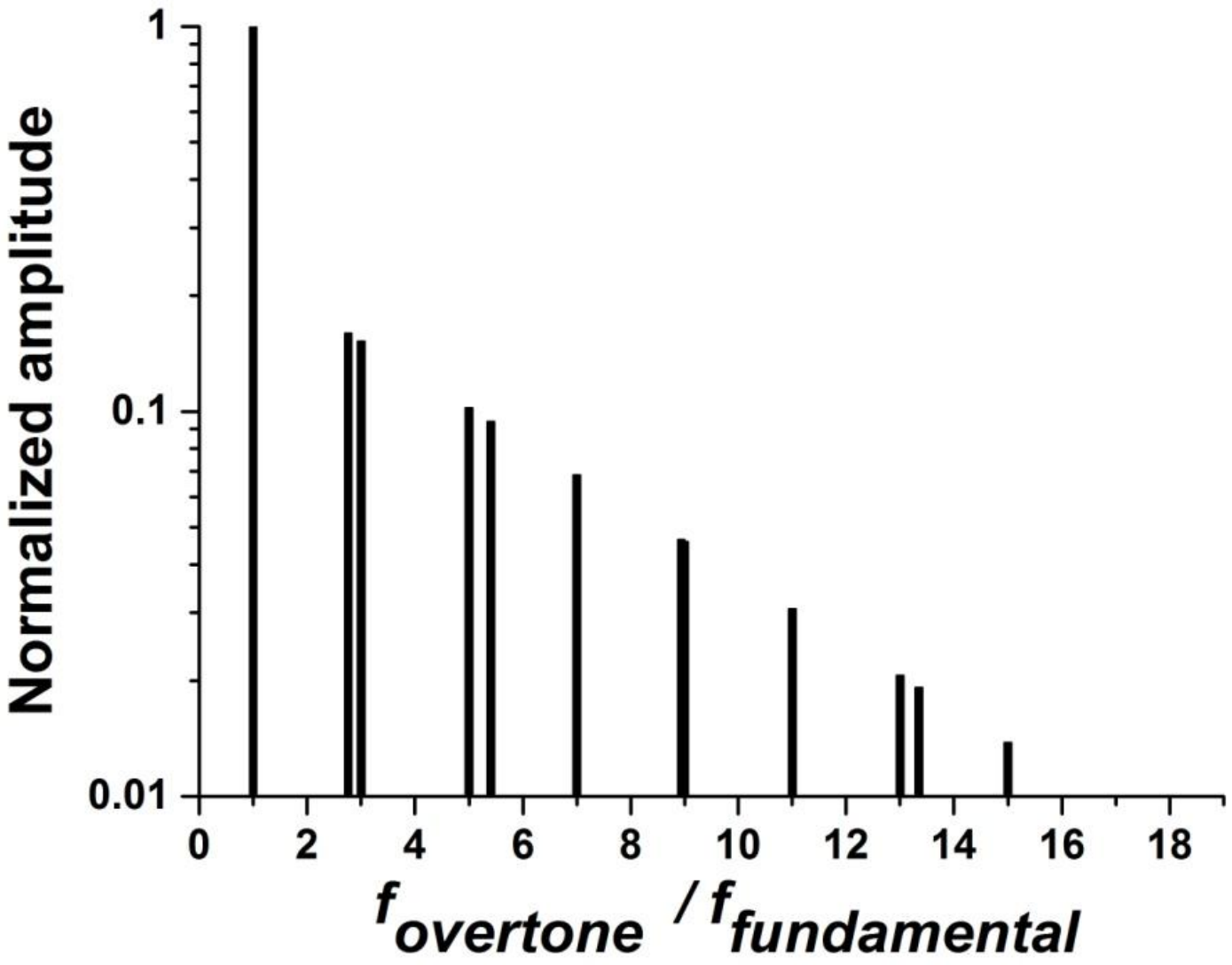

**Figure 4.** Spectrum corresponding to a rectangular bar free to vibrate at both ends with a cylindrical tubular resonator. The amplitude of the first overtone is taken to be 0.16 of the fundamental frequency, and the amplitude decays exponentially for the remaining successive overtones.

The dissonance curves produced using the three models (Sethares, 1993, 2005; Vassilakis, 2001) are shown in Figure 5. The three models predict a peak of maximum dissonance located in the range $[1.00, 1.15]$, and three minima of low dissonance at $\alpha = 1.67$, 1.81, and 1.96. From the second model of Sethares and from the Vassilakis model, the surrounding of the peaks $\alpha < 1.15$, $\alpha = 1.67$, and $\alpha = 1.81$ are the regions in which small variations in the $\alpha$ parameter produce the largest changes in the dissonance level.

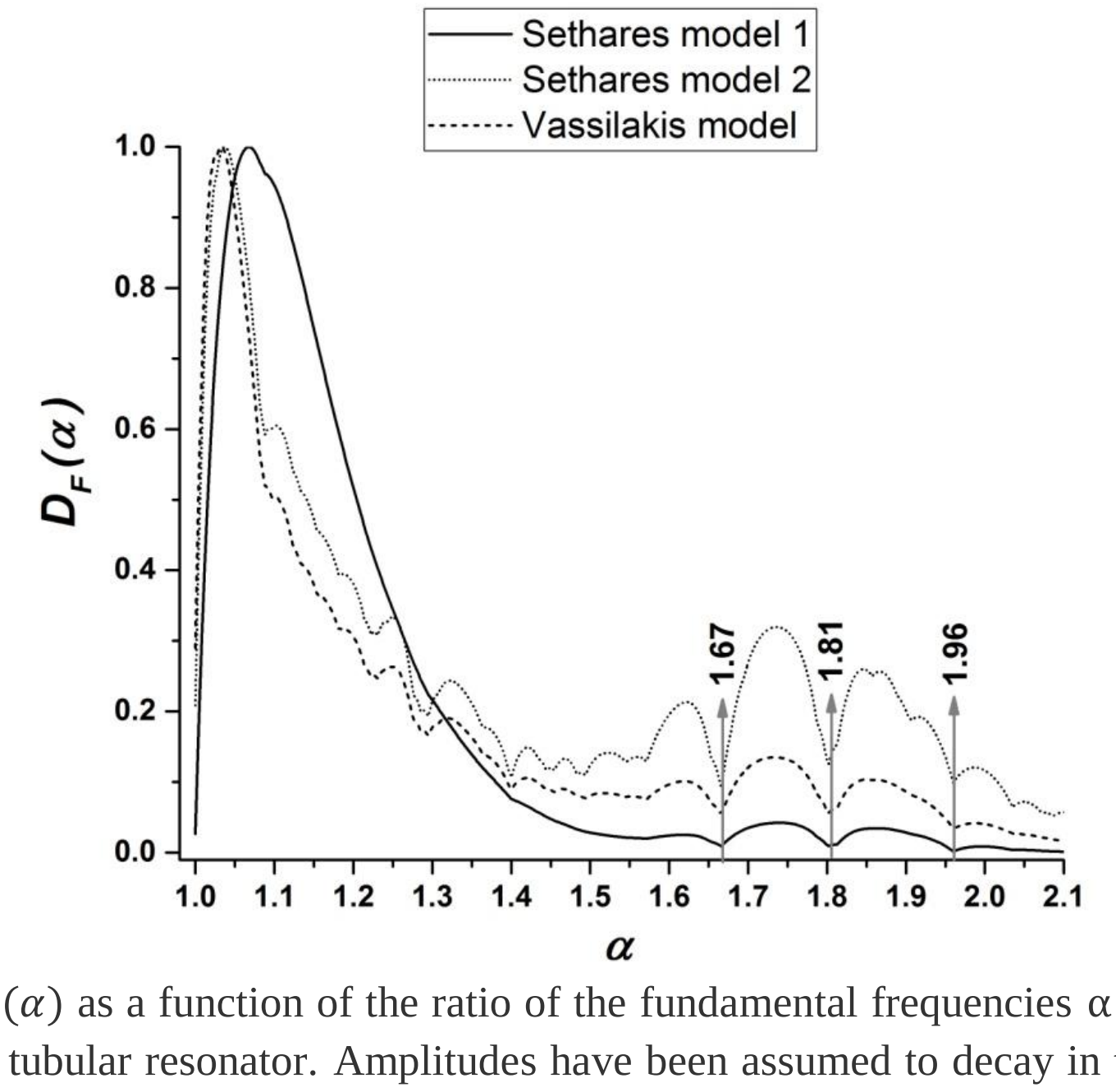

**Figure 5.** Dissonance level $D_F(\alpha)$ as a function of the ratio of the fundamental frequencies α for a rectangular bar free to vibrate at both ends with a cylindrical tubular resonator. Amplitudes have been assumed to decay in the form presented in Figure 4. The dissonance level scale has been normalized between 0 and 1. The lowest fundamental frequency is taken to be 300 Hz.

## Experimental dissonance level curves

The experimental analysis uses the recordings of the sound produced by each bar coupled with its resonator. The fundamental and first ten overtones for all marimbas were found; this is shown graphically for all bars of each marimba in Supplement 5. In most of the studied marimbas the fundamental has the largest amplitude; however, in the case of the Torres Family (Marimbas 9, 10, and 11 in Supplement 5) some overtones have larger amplitudes than the fundamental. Figure 6 shows, for each marimba, the average normalized amplitude as a function of the ratio between the corresponding overtone and the fundamental. The amplitude has been averaged over all the bars of a particular marimba.

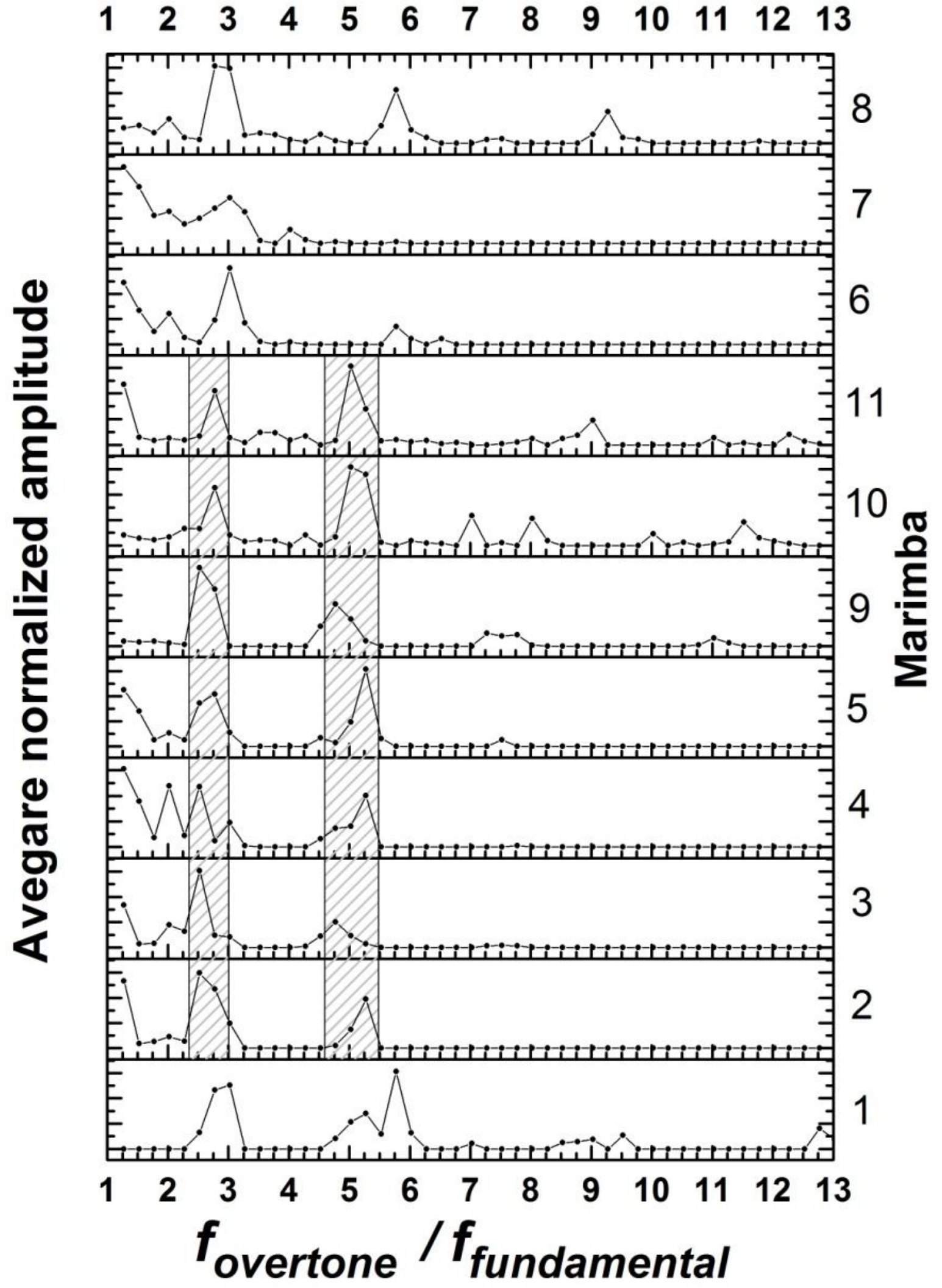

**Figure 6.** Average normalized amplitude for each marimba. The amplitudes have been averaged using a bin width of 0.25 for the ratio between the frequency of the overtones and the fundamental frequency.

Traditional marimbas 2, 3, 4, 5, 9, 10, and 11 exhibit a common pattern in the spectrum. Specifically, there are important incidences of overtones around $2.4 < n < 3.0$ and $4.6 < n < 5.5$. The normalized amplitudes obtained from superposition of the samples from all traditional marimbas are shown in Figure 7; the most important overtones are located near $n = 2.7 \pm 0.2$, $5.0 \pm 0.2$, and $5.4 \pm 0.2$. The values of these overtones are consistent with those of the theoretical model of the first and the second transverse modes of a rectangular bar free to vibrate at both ends and the second mode of a closed pipe. Other modes are absent or have low amplitudes in almost all cases, such as the unexpected peaks located at $n = 2.0$ that we associate with the first overtone of broken resonators functioning as open pipes.

The experimental values for the ratios of the heights of the peaks in Figure 7 are $A_1/A_0 \cong 0.16$, $A_2/A_0 \cong 0.13$, and $A_3/A_0 \cong 0.06$, where $A_0$, $A_1$, $A_2$, and $A_3$ are the amplitudes of the fundamental, first overtone, second overtone, and third overtone, respectively. These ratios of the amplitudes are approximately reproduced using the theoretical model proposed for the fundamental frequency and overtones $n = 2.758$, $5.000$, and $5.406$ (see Figure 4).

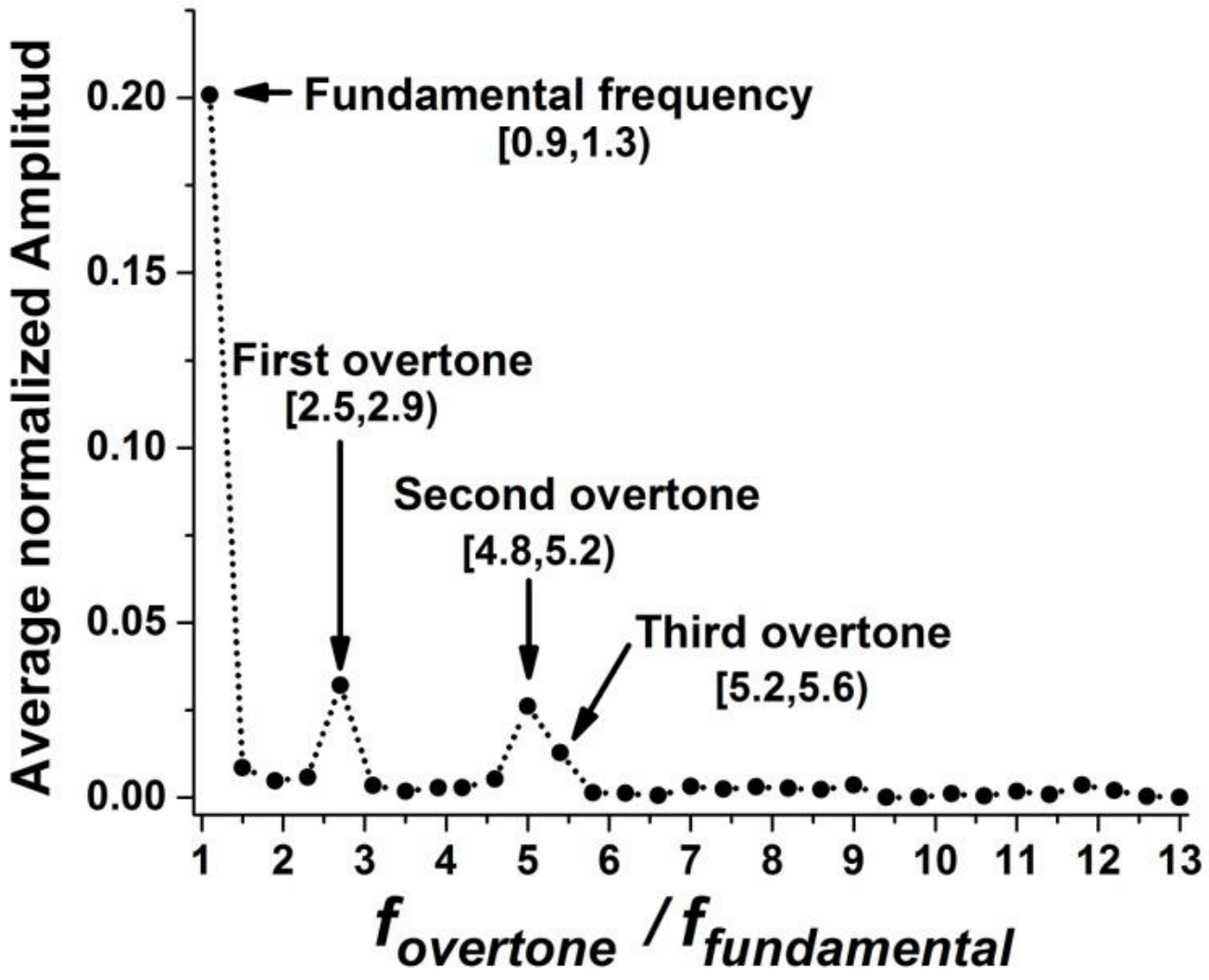

**Figure 7.** Average value of amplitudes associated with the overtones appearing in the traditional marimbas. Bin width of 0.4 for the ratio between the frequency of the overtones and the fundamental frequency.

Using overtones $n = 2.758$, $5.000$, and $5.406$ as representative values for the most important contributions found experimentally, and with the values of the normalized amplitudes presented previously, the corresponding dissonance curves were generated for both the Sethares and Vassilakis models (Figure 8a). The case of equal amplitudes is also presented; as in the low registers of the marimbas made by the Torres family, the amplitudes of the overtones are comparable to those of the fundamental (Supplement 5).

Comparing the experimental result (Figure 8a) with the theoretical model for a bar and its tubular resonator (Figure 5), illustrates that the dissonance curves have similar shapes with a large peak of dissonance, for $\alpha < 1.15$, and two narrow peaks of low dissonance, at $\alpha = 1.81$ and $\alpha = 1.96$. The most notable differences between the curves are smoothness, and the presence of a low dissonance narrow peak at $\alpha = 1.67$ in the theoretical model. Figure 8b shows the differentiation $dD_F(\alpha)/d\alpha$. There are two regions in which small variations in the $\alpha$ parameter produce large changes in the dissonance level $D_F(\alpha)$. These regions contain the musical intervals around the seconds, for $\alpha < 1.15$, and the sevenths, $\alpha \cong 1.81$, that are commonly avoided in the traditional marimba music.

The peak located at $\alpha \cong 1.67$ in the dissonance curve from the theoretical analysis of the bar coupled to the resonator (Figure 5) is due to the first overtone generated by the vibration of the tubular resonator $n$=3.0. For the most traditional marimbas this overtone presents insignificant values of amplitude, however for the marimbas 6, 7, and 8, this overtone contributes significantly (Figure 6). The ratio 1.67 is very close to the major sixths in the TTET scale ($\alpha \cong 1.68$); the statistical analysis finds that this interval occurs in only 2-3%, on average, of musical pieces (Table 6), indicating that the region around this peak is not commonly used in the *marimba de chonta* music.

The peak located at $\alpha \cong 1.96$ is due to the overtone $n = 5.41$. Since the amplitude for this overtone is smaller than for overtones $n = 2.76$, and $n = 5.00$, the dissonance curve changes less abruptly in the region of the octaves, when compared to those in the region of the sevenths. Perhaps this is the reason the octaves are not suppressed in the geometrical schemes of Figure 1.

## The transposition practice

The transposition of a musical piece involves rewriting a composition in a different key, a practice that is common when accommodating different voice ranges (Apel, 1974). Since in an equal tempered scale the frequency ratios are preserved, the dissonance sequence in a transposition process remains almost unchanged conserving similar tension-relaxation sequences.

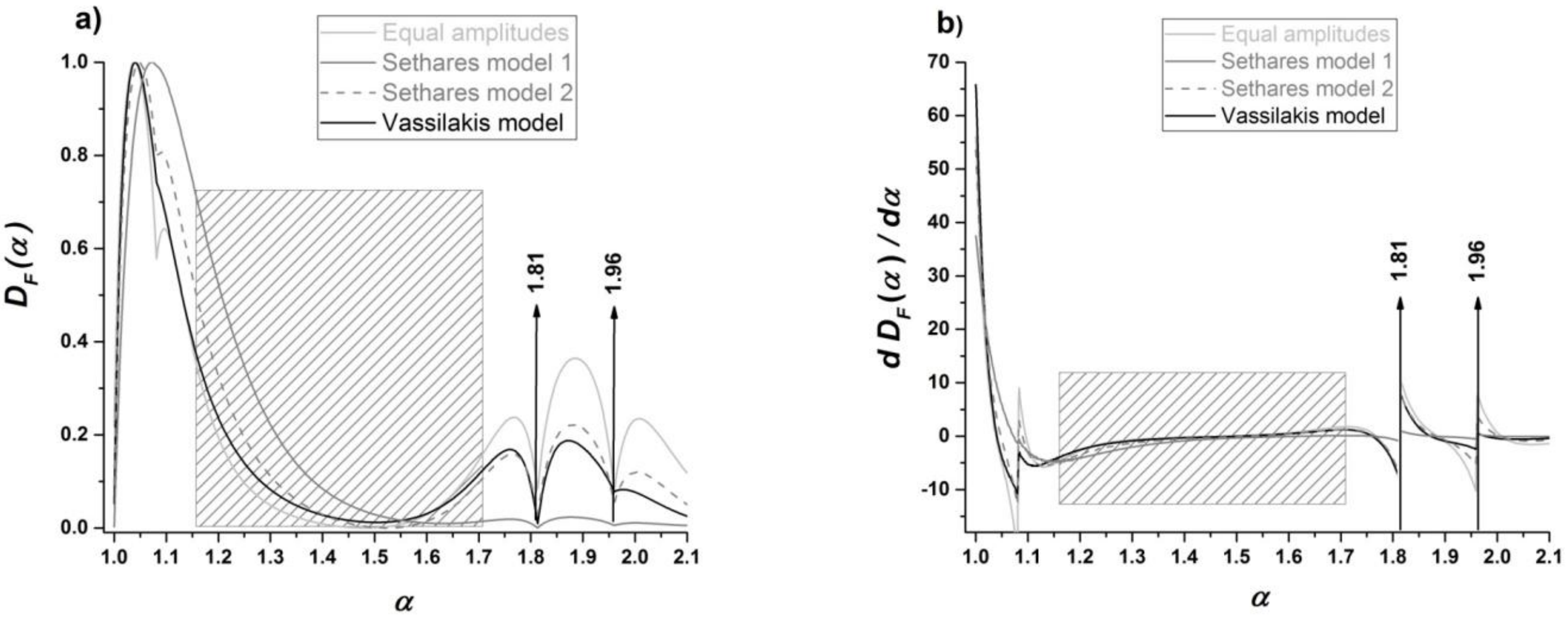

**Figure 8.** a) Dissonance level $D_F(\alpha)$ as a function of the ratio of the fundamental frequencies $\alpha$ for the most important components of the experimental spectrum. b) Differentiation of the dissonance level with respect to the fundamental frequency ratio $dD_F(\alpha)/d\alpha$ as a function of the fundamental frequency ratio $\alpha$, for the most important components of the experimental spectrum. The dissonance level scale has been normalized between 0 and 1. The lowest fundamental frequency has been taken to be 300 Hz.

In the isotonic scales, the transposition process keeps the same frequency ratios for the same geometrical distances. In traditional marimbas the average frequency ratios are the ones that follow isotonic scales, with relatively small deviations with respect to the mean (see Tables 2, 3, 4, 5), suggesting that similar successive dissonance changes can be preserved if the regions with large changes in dissonance are avoided.

Up to know, the dissonance curves have been constructed over a fixed value for one of the fundamental frequencies ($300\,Hz$). In general, the form of the dissonance curves is similar for different values of the fixed fundamental frequency. For example, Figure 9a shows the dissonance curves for different fixed fundamental frequencies using the second model of Sethares using the spectrum found experimentally. The set of fundamental frequencies was obtained using the average ratios of the fundamental frequencies found for adjacent bars of the equi-heptatonic traditional marimbas of the present study. Hence this figure represents the dissonance curves starting from different bars of the marimba. This figure also shows that the same musical interval played in a lower part of the register tends to be more dissonant than when played in a higher part, a well-known property of musical intervals (Plomp & Levelt, 1965; Roederer, 2009).

In the performance of a musical piece on a traditional marimba different simultaneous pairs of bars are hit to construct successive dissonance levels. Since the melodic motion is produced using bars near to each other (Miñana, 2010a), this practice results in jumps between different dissonance curves near to each other (Figure 9b). In the region of the dissonance curves of traditional marimbas that excludes the seconds, the sevenths, and the octaves, transitions or jumps made between near or adjacent dissonance curves lead to small changes in the dissonance sequences. For example, Figure 9b shows that the most dissonant elements in this music are usually due to intervals with a distance of 2 steps, and the most consonant are usually due to intervals with distances of 4 or 5 steps (in the equi-heptatonic scale). These features do not change significantly due to jumps between near or adjacent dissonance curves. A different situation occurs for 6, 7, and 8 step distances in the equi-heptatonic, equi-octatonic, and equi-enneatonic scales, respectively, in which narrow peaks can lead to changes from consonant to dissonant values for small changes in $\alpha$, or vice-versa, especially in the region of sevenths.

Regarding the transposition process, the bar used to start the performance of a piece determines the initial dissonance curve for the sequence of steps. If two dissonance curves are adjacent or close to each other, then the same sequence of harmonic intervals generated using distances of 2, 3, 4, and 5 steps in the case of the equi-heptatonic marimbas (2, 3, 4, 5, and 6 for equi-octatonic, 2, 3, 4, 5, 6, and 7 for the equi-enneatonic) will produce similar sequences of dissonance changes, independently of which of the dissonance curves is the initial one. On the other hand, avoiding intervals in the regions with large variations in the dissonance level (1, 6, and 7 distance steps for equi-heptatonic marimbas; 1, 7, and 8 for equi-octatonic; 1,8, and 9 for equi-enneatonic) allows the preservation of geometrical distances of the sequence of intervals in the transposition process.

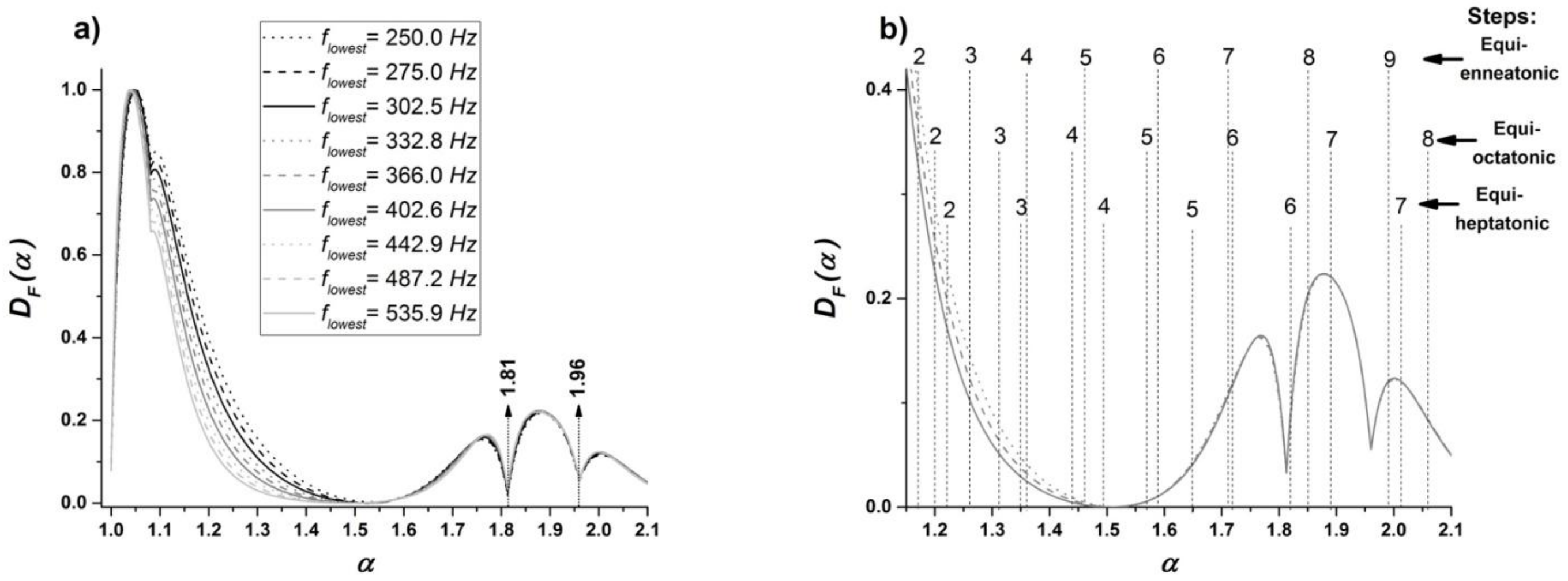

**Figure 9.** a) Dependence of the dissonance level $D_F(\alpha)$ with the lowest fundamental frequency for the most important components of the experimental spectrum for the case of exponentially decaying amplitudes. The dissonance level scale has been normalized between 0 and 1. b) Close up of the region used in the *marimba de chonta* music for three successive bars ($f_{lowest} = 332.8\ Hz, 366.0\ Hz, 402.6\ Hz$). Vertical dash lines represent the average tuning as a function of the distance between bars in steps in the case of the equi-heptatonic traditional marimbas {2, 3, 4, 5}, the equi-octatonic marimbas {9, 10}, and the equi-enneatonic marimba 11.

## Conclusions

We analyzed the theoretical and experimental spectra of the *marimba de chonta* and the tuning of this instrument using data collected in 2015. We found that tunings follow equi-heptatonic, equi-octatonic and equi-enneatonic scales with relatively small uncertainties with respect to the precise values that would result from the mathematical relation associated with isotonic scales. A comparison with a previous study carried out using data collected in the 80's indicates that the tuning of traditional marimbas constructed with equi-heptatonic scales has since changed from low octaves to almost just octaves. This mathematical analysis, together with the testimonies of musicians and instrument makers, suggests that this change is due to the influence of Western music.

One of the main approaches for understanding tuning is based on the acoustical properties of the musical instrument; the tuning must be done in the positions of the local minima of dissonance generated by the spectrum of the timbre of the musical instrument, independent of the musical practices. We found that the tunings of traditional marimbas approximately follow this principle, as the most used harmonic intervals are placed in a broad minimum of dissonance. The main relevance of this broad minimum of dissonance is the possibility of playing at the same time bars separated by a specific geometric distance in regions close to each others in the register, while keeping similar levels of dissonance. This feature also allows for the transposition of full arrays of intervals, while approximately maintaining the main harmonic and melodic information in the tension-relaxation sequences. However, two sharp minima of dissonance characterized by narrow peaks are avoided, as they limit the transposition practice. This is because small uncertainties in tuning lead to large variations in dissonance of intervals of the same distance, altering the tension-relaxation sequences. This rational is reinforced by the suppression of intervals close to the seconds, as they are placed in a region with a narrow maximum of dissonance.

In the case of tunings with large changes in the dissonance level for small variations in the tuning, as in the case of narrow peaks of dissonance, any uncertainty in the tuning limits the possibility of practicing any transposition principle. Vice versa, tunings with intrinsic uncertainties require broad soft minima of dissonance; otherwise any transposition practice is avoided.

Finally, the interplay between three different attributes of musical phenomena—consonance, tuning, and musical practices—connecting the microscopic and mesoscopic characteristics of music, opens new possibilities for understanding music in the realm of complexity.

## Acknowledgments

We thank Universidad Nacional de Colombia for funding this study under grant HERMES 19010. We thank Carlos Miñana, Hector Tascón, Yeiner Orobio and Julián Villegas for their valuable comments and suggestions. We also thank Roberto García, Maria Angélica Bejarano, Jesús Guevara and Alejandro Pernet for the recording of the marimba samples. We thank Kevin Pineda for the transcription of musical scores. Finally, we thank the many *marimba de chonta* instrument makers, musicians, and singers that we met in the Pacific Coast of Colombia.

# Supplement 1

# Marimba de Chonta tunings
# (Carlos Miñana)

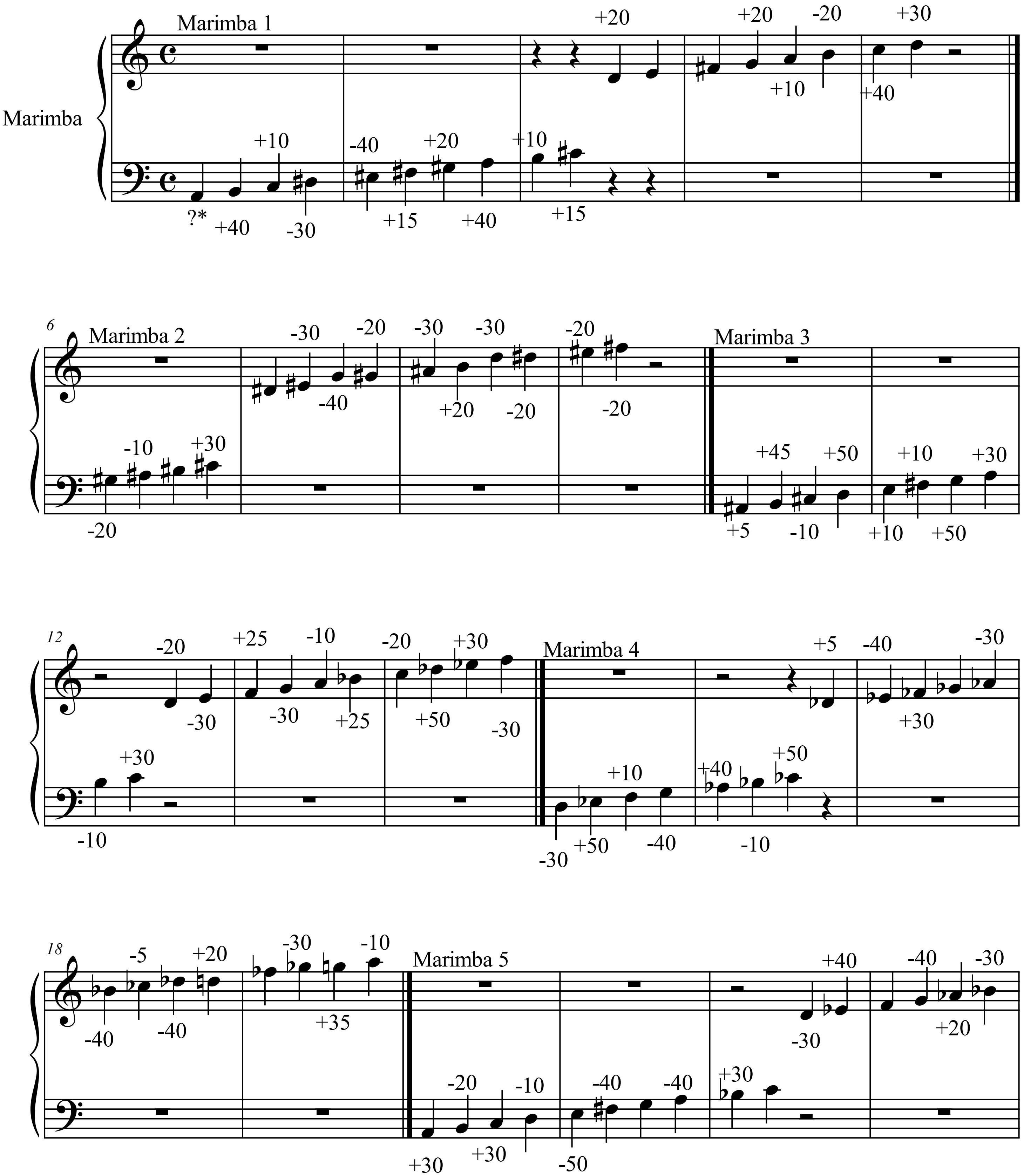

* In the analysis we have used the pitch A without taking into account the symbol "?"

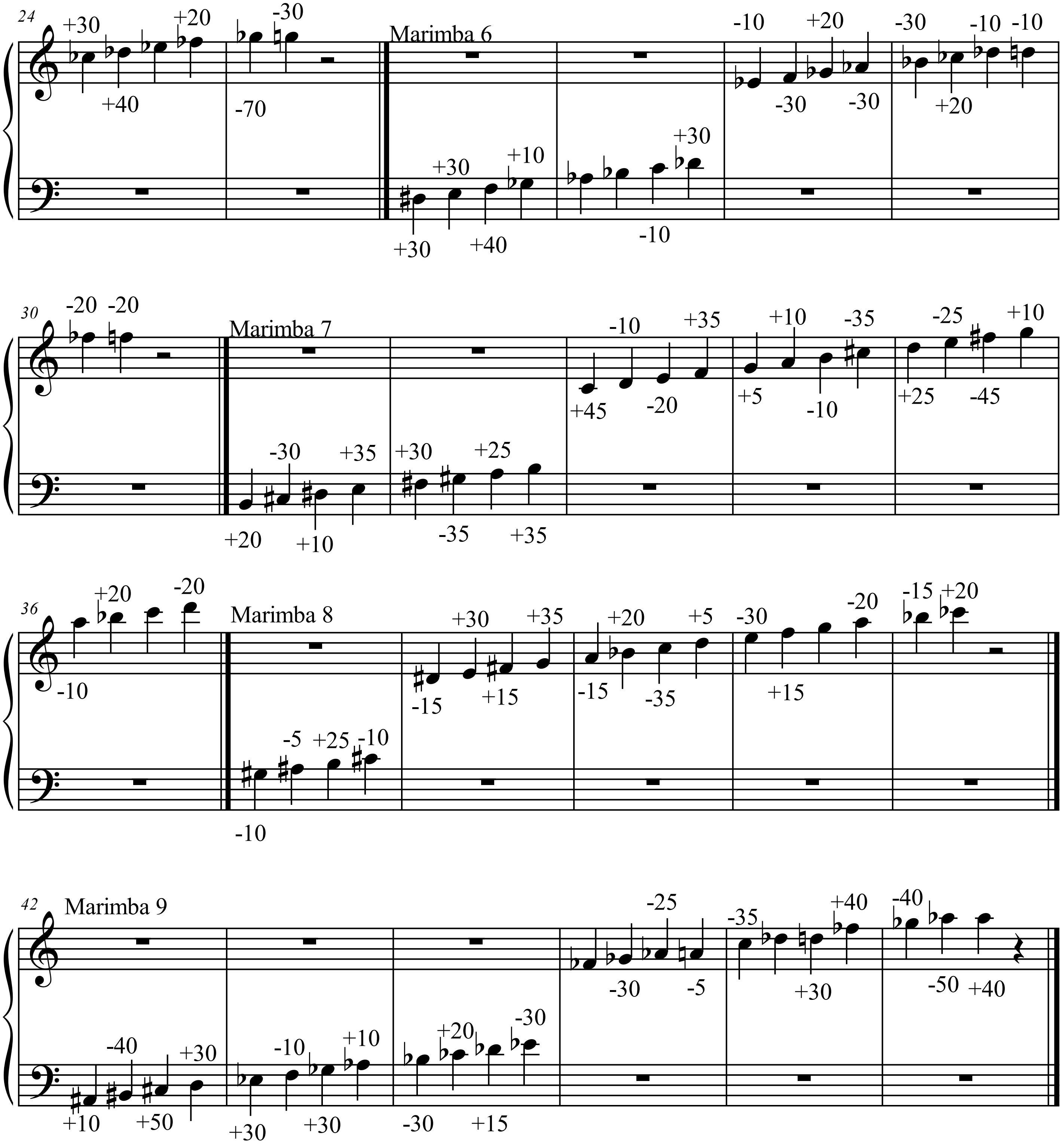

# Supplement 2

**Fundamental frequencies produce by each bar with its corresponding resonator.**

The uncertainty values correspond to the distance between adjacent frequencies in the FFT analysis assuming that it corresponds to the uncertainty in the peak location. The mark “*” refers to unconfident values for the fundamental frequencies.

| | | Fundamental frequency (*Hz*) | | | | | | | | | | |
|---|---|---|---|---|---|---|---|---|---|---|---|---|
| | | **Marimba 1** | **Marimba 2** | **Marimba 3** | **Marimba 4** | **Marimba 5** | **Marimba 6** | **Marimba 7** | **Marimba 8** | **Marimba 9** | **Marimba 10** | **Marimba 11** |
| | **1** | 131.62 ± 1.19 | 270.85 ± 1.96 | 184.70 ± 0.95 | 167.91 ± 1.55 | 158.66 ± 1.10 | 171.93 ± 0.75 | 190.21 ± 3.59 | 145.71 ± 1.58 | 191.01 ± 3.35 | 130.00 ± 9.35 * | 169.39 ± 7.70 * |
| | **2** | 145.93 ± 1.27 | 305.01 ± 2.52 | 204.74 ± 0.85 | 188.53 ± 1.61 | 170.82 ± 1.09 | 175.40 ± 1.16 | 210.60 ± 2.70 | 167.24 ± 1.61 | 205.53 ± 2.23 | 123.35 ± 5.36 | 143.81 ± 4.11 * |
| | **3** | 165.23 ± 1.09 | 341.11 ± 1.71 | 223.41 ± 1.16 | 209.54 ± 1.64 | 190.35 ± 1.63 | 200.89 ± 0.81 | 226.28 ± 2.76 | 180.60 ± 1.67 | 227.84 ± 2.45 | 128.76 ± 6.78 | 209.20 ± 7.21 * |
| | **4** | 175.38 ± 1.07 | 367.60 ± 2.12 | 251.43 ± 1.30 | 219.15 ± 2.15 | 214.15 ± 1.38 | 221.17 ± 1.14 | 249.61 ± 3.12 | 195.12 ± 1.68 | 247.56 ± 3.26 | 148.30 ± 6.18 | 153.34 ± 4.38 |
| | **5** | 197.31 ± 1.01 | 409.85 ± 2.55 | 282.56 ± 2.05 | 241.62 ± 2.95 | 229.48 ± 2.64 | 250.50 ± 1.06 | 270.62 ± 2.73 | 207.33 ± 1.69 | 272.75 ± 1.98 | 147.50 ± 3.51 | 163.50 ± 5.64 |
| | **6** | 220.47 ± 0.87 | 450.50 ± 2.20 | 309.48 ± 2.36 | 275.01 ± 1.72 | 259.31 ± 1.81 | 267.41 ± 0.96 | 305.05 ± 3.32 | 229.89 ± 1.70 | 287.25 ± 4.29 | 169.88 ± 3.15 | 184.31 ± 4.98 |
| | **7** | 247.30 ± 0.87 | 496.49 ± 2.96 | 334.65 ± 1.95 | 298.57 ± 2.16 | 294.34 ± 3.16 | 292.42 ± 1.40 | 332.30 ± 2.66 | 252.85 ± 1.72 | 313.55 ± 5.50 | 187.36 ± 2.80 | 198.21 ± 2.75 |
| | **8** | 261.68 ± 0.81 | 536.73 ± 3.21 | 367.40 ± 1.90 | 331.03 ± 3.09 | 316.90 ± 2.08 | 334.85 ± 1.50 | 365.02 ± 4.80 | 285.65 ± 1.75 | 349.14 ± 3.49 | 207.35 ± 1.13 | 207.13 ± 1.59 |
| | **9** | 293.58 ± 1.17 | 609.06 ± 2.11 | 413.15 ± 3.36 | 369.93 ± 3.36 | 349.58 ± 1.42 | 360.21 ± 1.31 | 396.51 ± 2.83 | 315.69 ± 1.82 | 379.01 ± 5.83 | 219.98 ± 1.86 | 215.10 ± 2.95 |
| | **10** | 330.00 ± 0.91 | 687.68 ± 3.10 | 451.49 ± 2.57 | 413.35 ± 2.48 | 394.01 ± 1.70 | 387.14 ± 1.18 | 449.56 ± 2.71 | 355.90 ± 1.70 | 409.60 ± 4.82 | 243.93 ± 1.16 | 233.14 ± 3.33 |
| | **11** | 351.58 ± 1.03 | 747.17 ± 2.20 | 500.13 ± 2.10 | 440.39 ± 2.79 | 430.19 ± 1.67 | 438.80 ± 1.31 | 479.41 ± 3.05 | 390.45 ± 1.72 | 462.12 ± 4.32 | 262.04 ± 1.48 | 263.92 ± 5.50 |
| **Bar** | **12** | 393.69 ± 1.02 | 818.48 ± 2.14 | 542.98 ± 2.64 | 486.81 ± 4.13 | 472.91 ± 1.86 | 481.43 ± 1.23 | 543.40 ± 3.25 | 429.08 ± 1.88 | 495.17 ± 7.99 | 294.23 ± 1.56 | 283.19 ± 10.11 |
| | **13** | 439.43 ± 0.99 | 910.90 ± 2.45 | 634.07 ± 2.59 | 539.29 ± 2.84 | 535.00 ± 2.83 | 518.89 ± 1.47 | 600.83 ± 2.78 | 477.37 ± 1.80 | 539.33 ± 4.86 | 311.65 ± 3.21 | 299.34 ± 3.29 |
| | **14** | 495.44 ± 0.94 | 994.77 ± 3.06 | 672.32 ± 3.14 | 585.53 ± 3.33 | 582.64 ± 2.03 | 580.70 ± 1.66 | 655.81 ± 3.49 | 515.07 ± 1.73 | 603.41 ± 4.08 | 356.91 ± 1.53 | 312.88 ± 2.87 |
| | **15** | 523.24 ± 1.14 | 1080.37 ± 2.60 | 740.70 ± 3.53 | 644.27 ± 3.64 | 650.65 ± 2.85 | 654.80 ± 1.87 | 719.59 ± 5.00 | 572.42 ± 1.73 | 636.04 ± 5.30 | 381.38 ± 1.94 | 358.66 ± 3.04 |
| | **16** | 588.65 ± 1.20 | 1240.32 ± 2.38 | 838.27 ± 3.95 | 737.47 ± 3.60 | 702.45 ± 3.07 | 705.79 ± 2.20 | 800.23 ± 2.91 | 648.36 ± 1.79 | 722.67 ± 6.07 | 428.30 ± 2.75 | 378.97 ± 1.81 |
| | **17** | 660.86 ± 1.08 | ------ | 896.38 ± 4.55 | 813.44 ± 3.48 | 798.79 ± 3.01 | 783.34 ± 2.12 | 910.42 ± 3.00 | 722.11 ± 1.83 | 755.61 ± 4.66 | 475.78 ± 2.07 | 412.29 ± 1.65 |
| | **18** | 700.72 ± 0.90 | ------ | 989.25 ± 3.41 | 876.62 ± 3.97 | 871.50 ± 4.49 | 859.39 ± 2.46 | ------ | ------ | 874.57 ± 5.95 | 528.51 ± 2.36 | 431.40 ± 3.17 |
| | **19** | 791.14 ± 0.84 | ------ | 1128.03 ± 4.27 | 986.17 ± 6.45 | ------ | 928.18 ± 2.37 | ------ | ------ | 935.74 ± 4.70 | 563.27 ± 2.40 | 481.15 ± 3.39 |
| | **20** | 878.25 ± 0.80 | ------ | 1248.11 ± 6.60 | 1131.23 ± 4.83 | ------ | 1011.77 ± 2.68 | ------ | ------ | 1057.87 ± 4.16 | 608.65 ± 2.32 | 527.56 ± 3.77 |
| | **21** | 985.22 ± 0.99 | ------ | 1342.93 ± 6.19 | 1224.65 ± 6.45 | ------ | ------ | ------ | ------ | 1121.73 ± 6.34 | ------ | 556.13 ± 4.56 |
| | **22** | 1044.23 ± 1.08 | ------ | 1448.71 ± 6.59 | 1338.57 ± 6.73 | ------ | ------ | ------ | ------ | ------ | ------ | 609.91 ± 5.08 |
| | **23** | ------ | ------ | 1663.24 ± 8.07 | 1472.77 ± 8.71 | ------ | ------ | ------ | ------ | ------ | ------ | 657.09 ± 4.60 |
| | **24** | ------ | ------ | 1767.66 ± 9.16 | 1643.50 ± 9.67 | ------ | ------ | ------ | ------ | ------ | ------ | 716.08 ± 4.56 |

# Supplement 3

| Musical piece number | Musician |
| --- | --- |
| 1 | Dioselino Rodríguez |
| 2 | |
| 3 | |
| 4 | |
| 5 | |
| 6 | Genaro Torres |
| 7 | Francisco Torres |

# Musical piece # 1

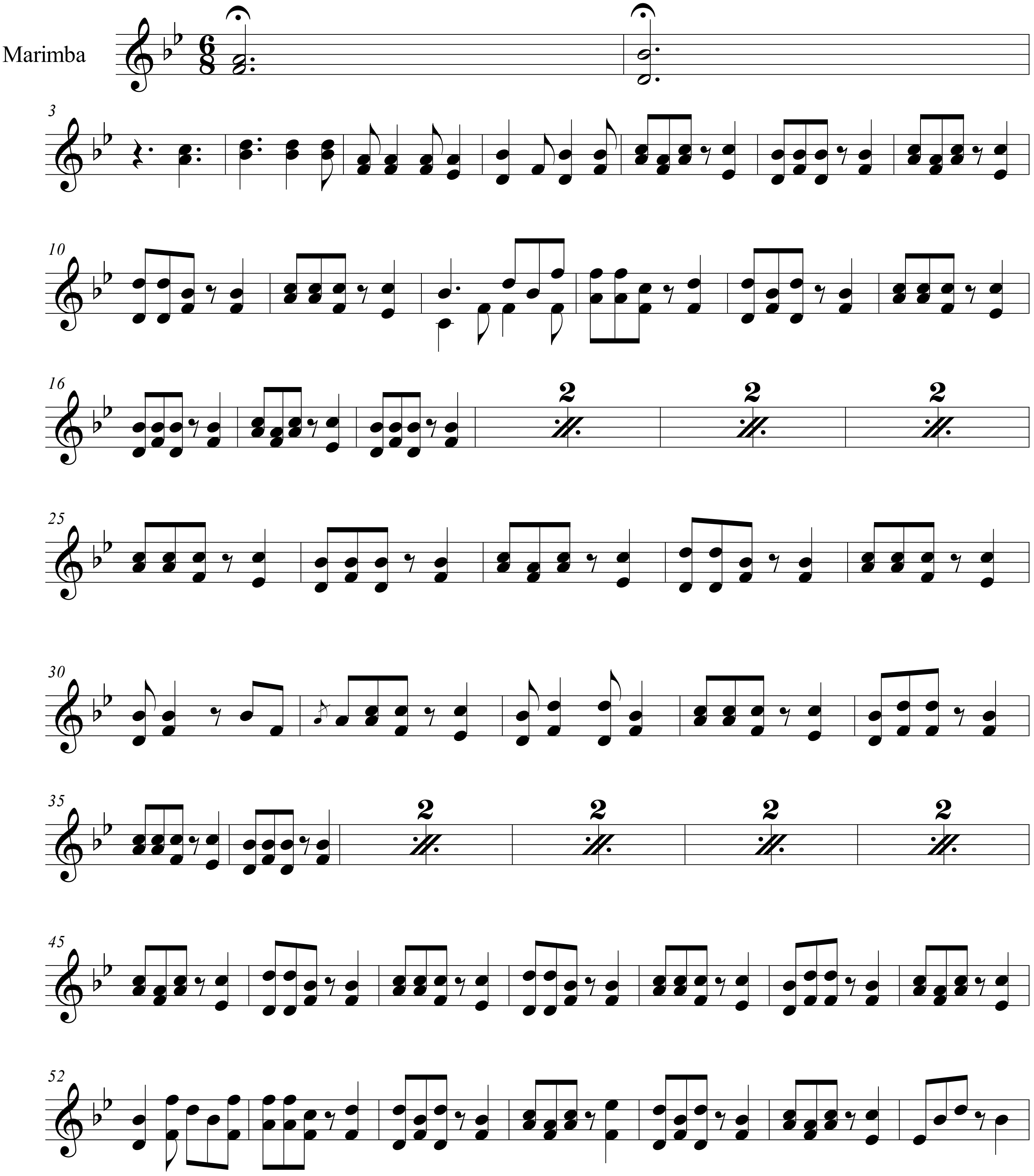

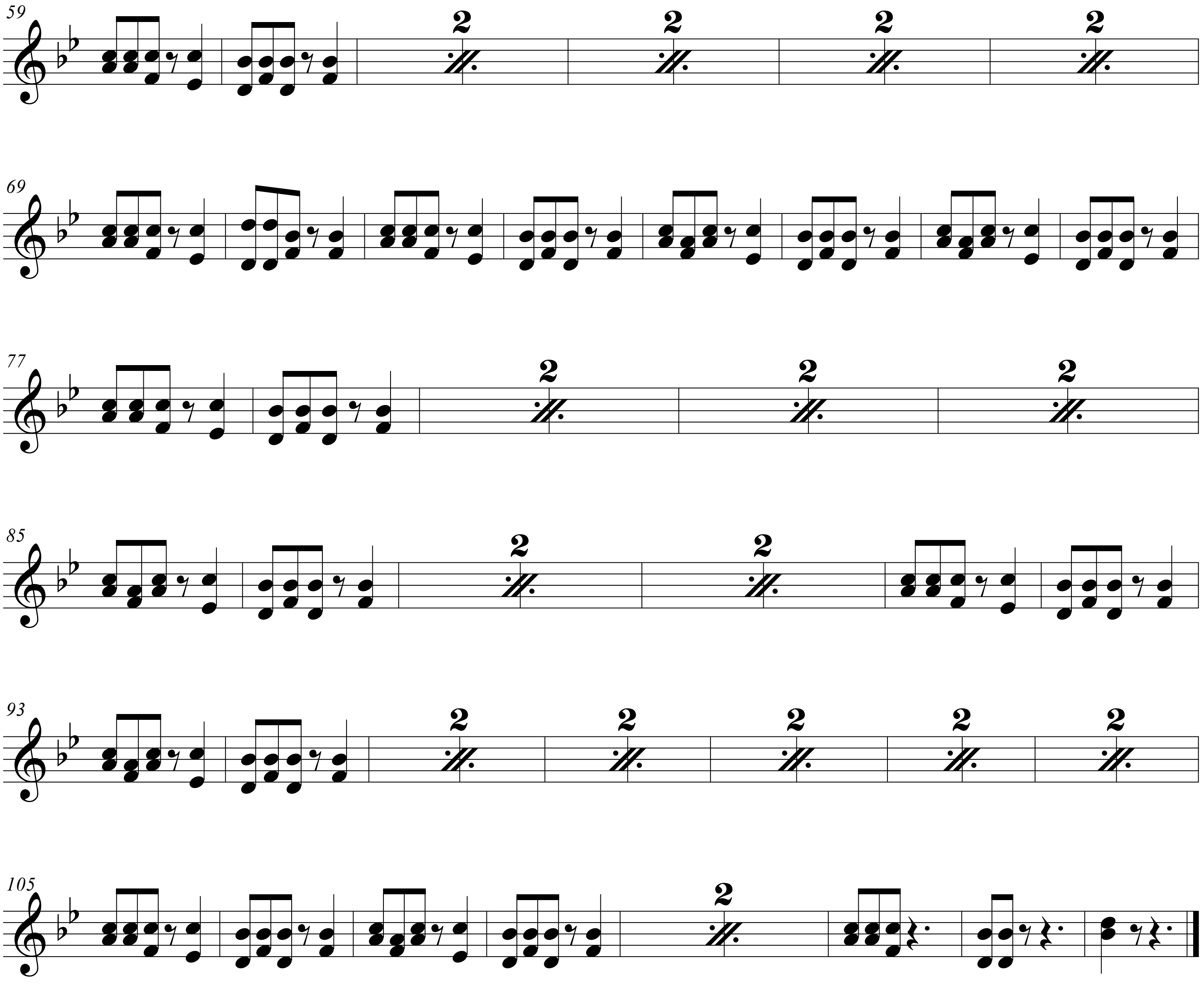

# Musical piece # 2

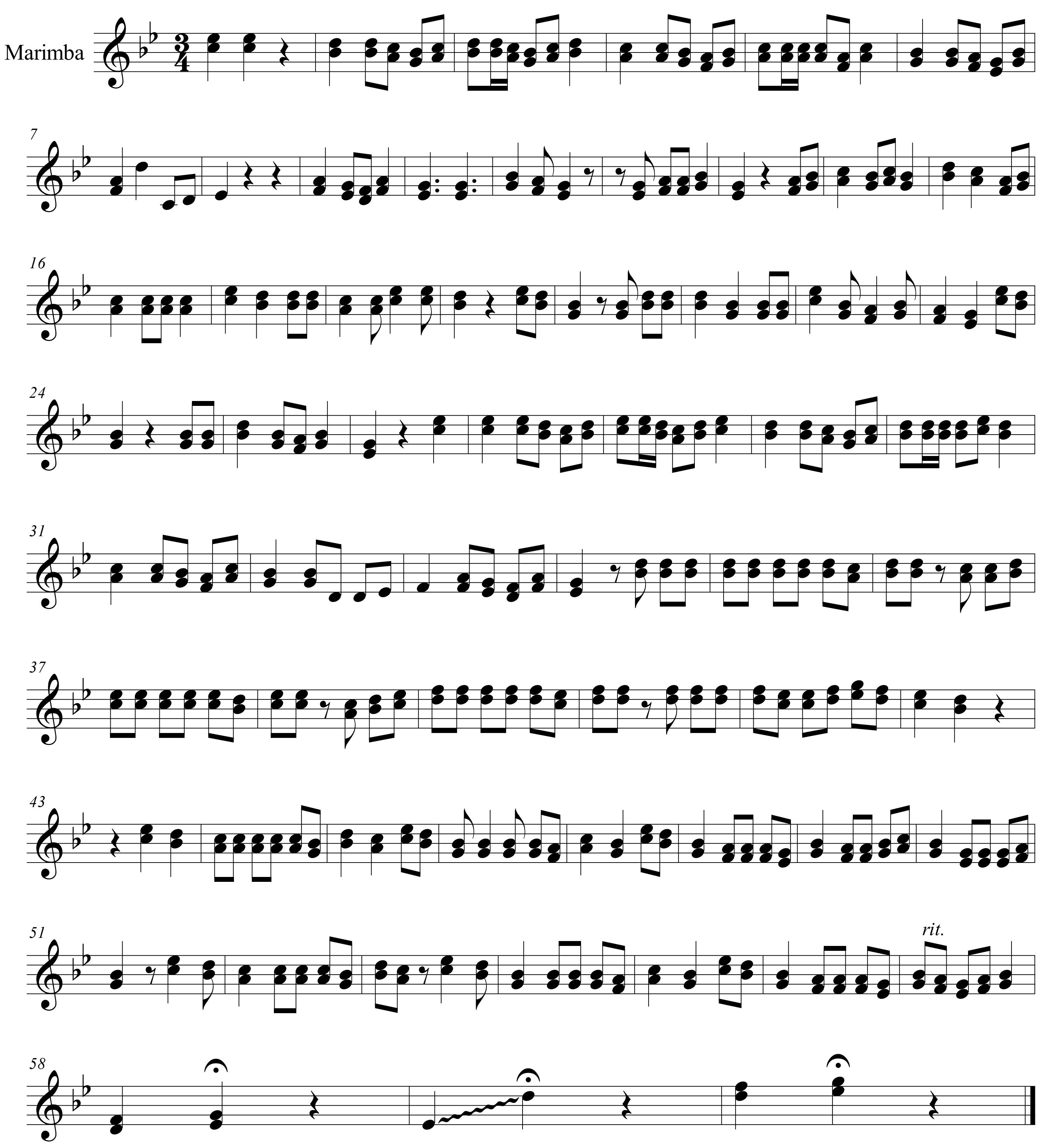

# Musical piece # 3

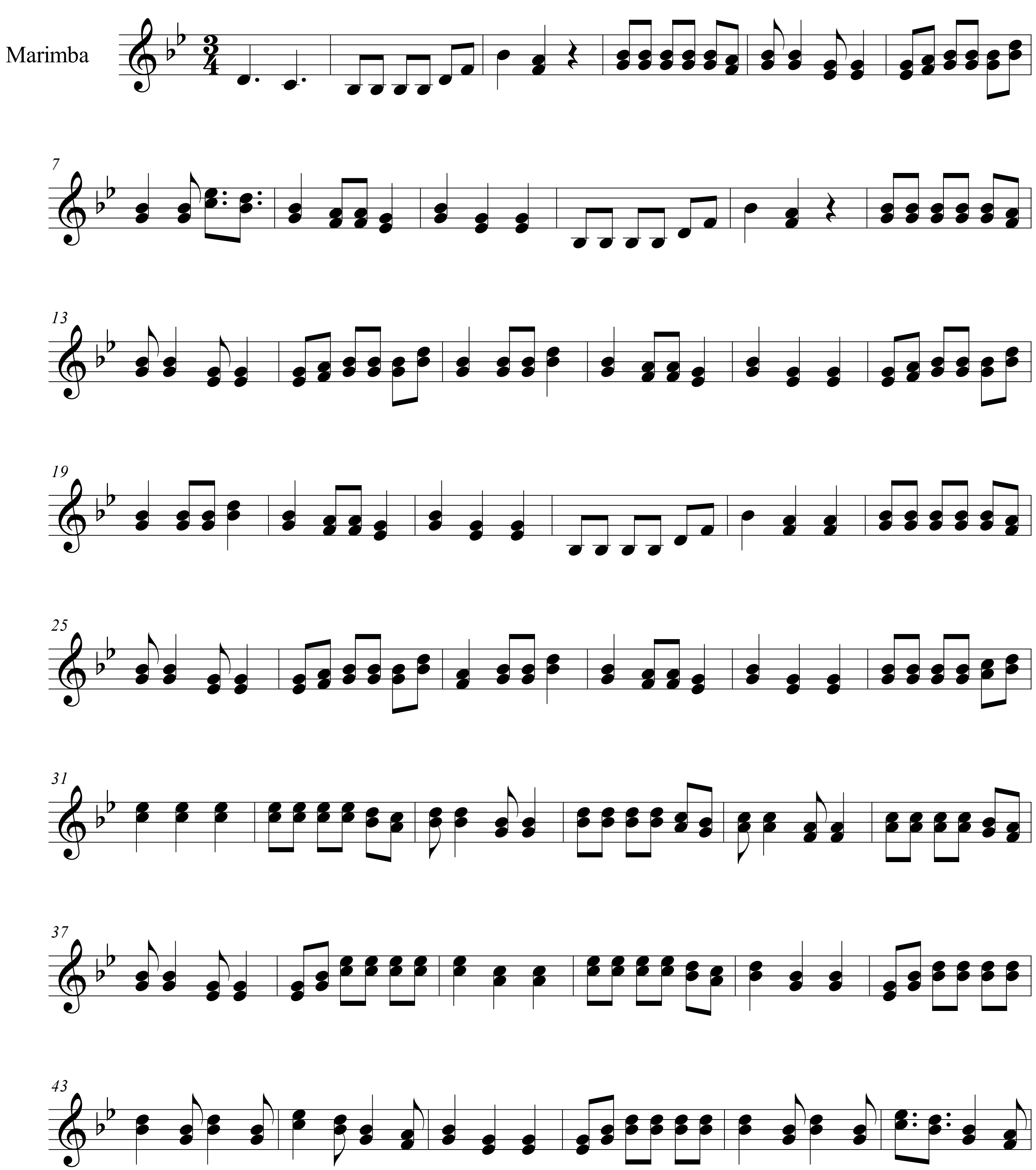

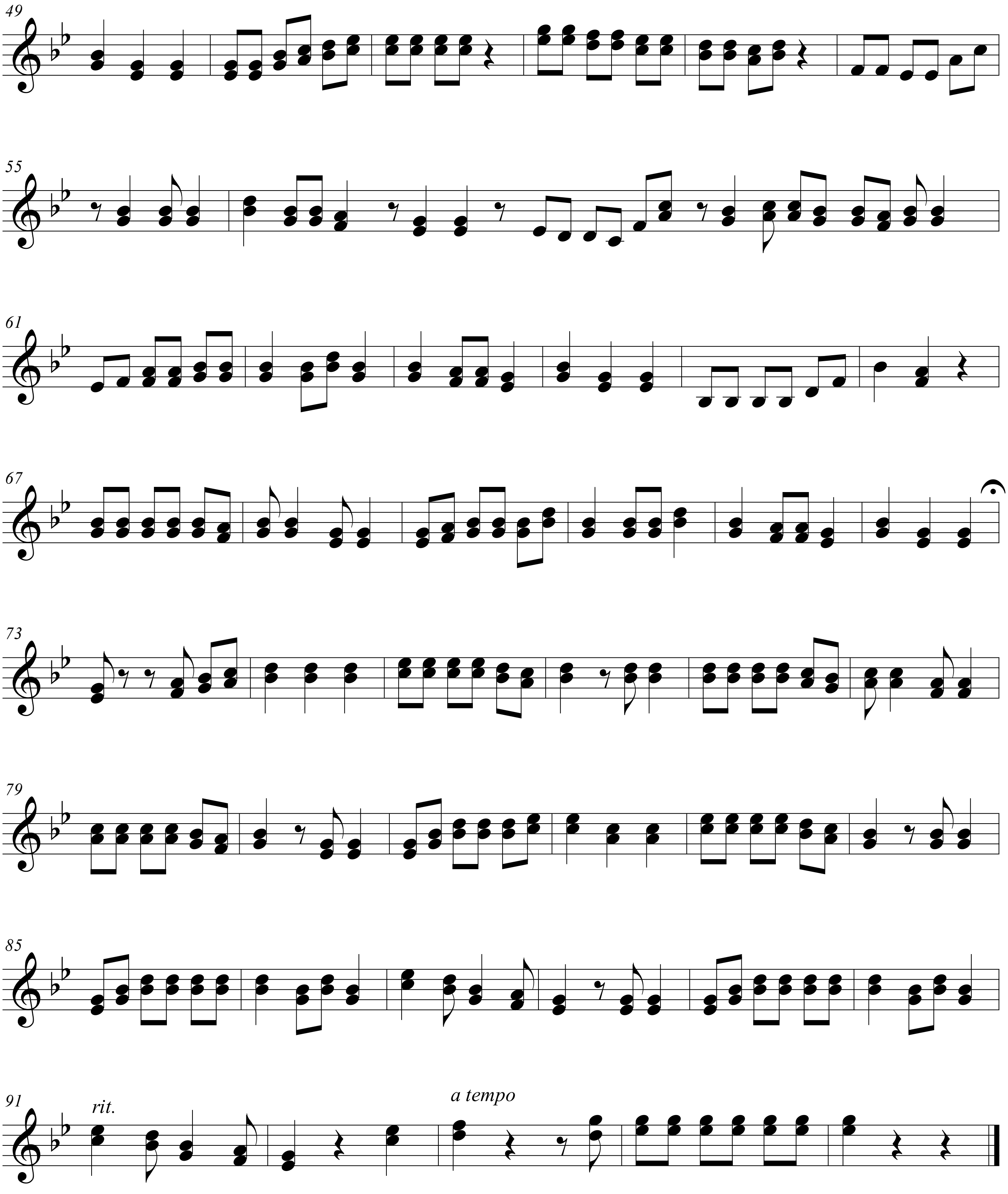

# Musical piece # 4

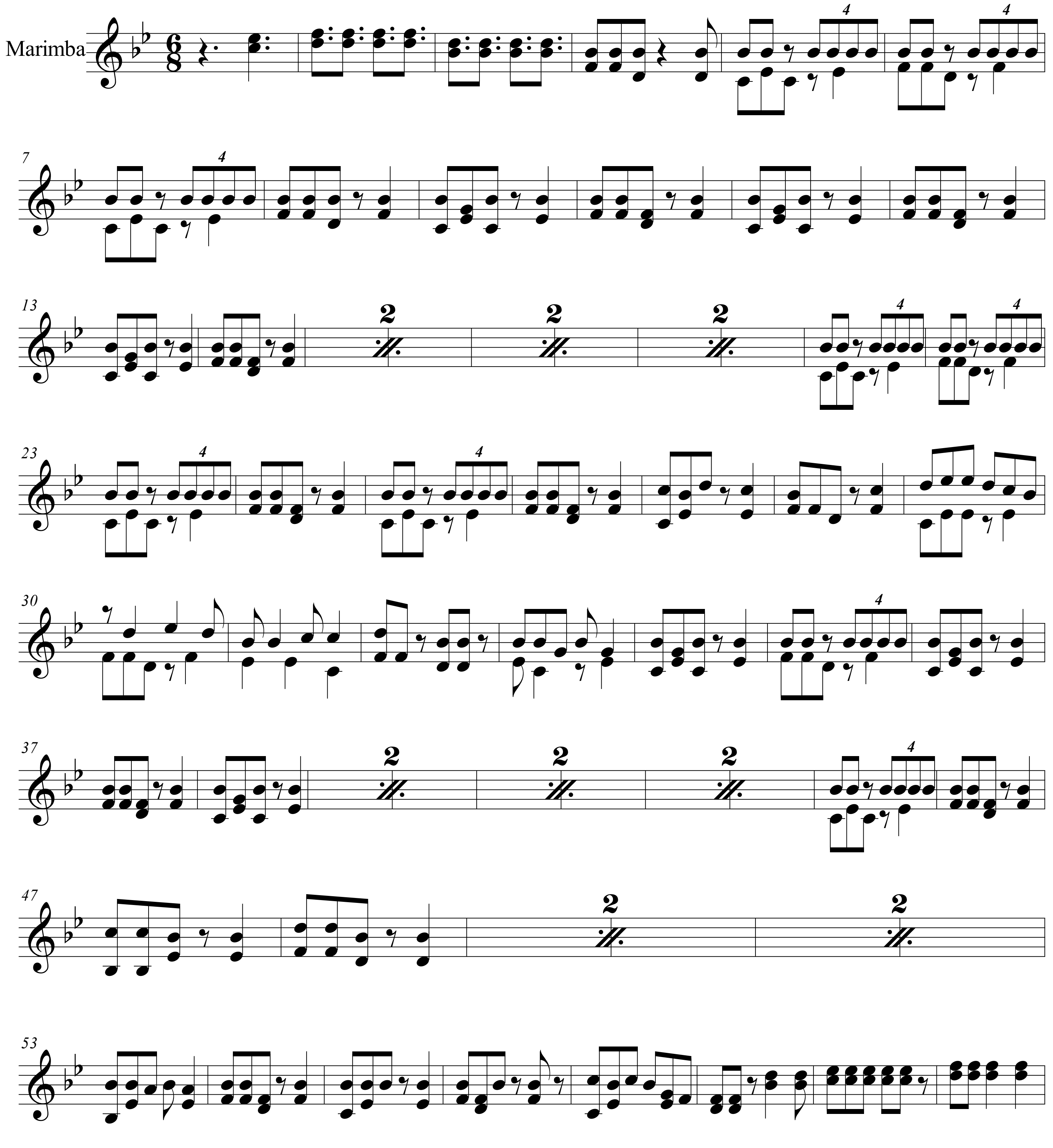

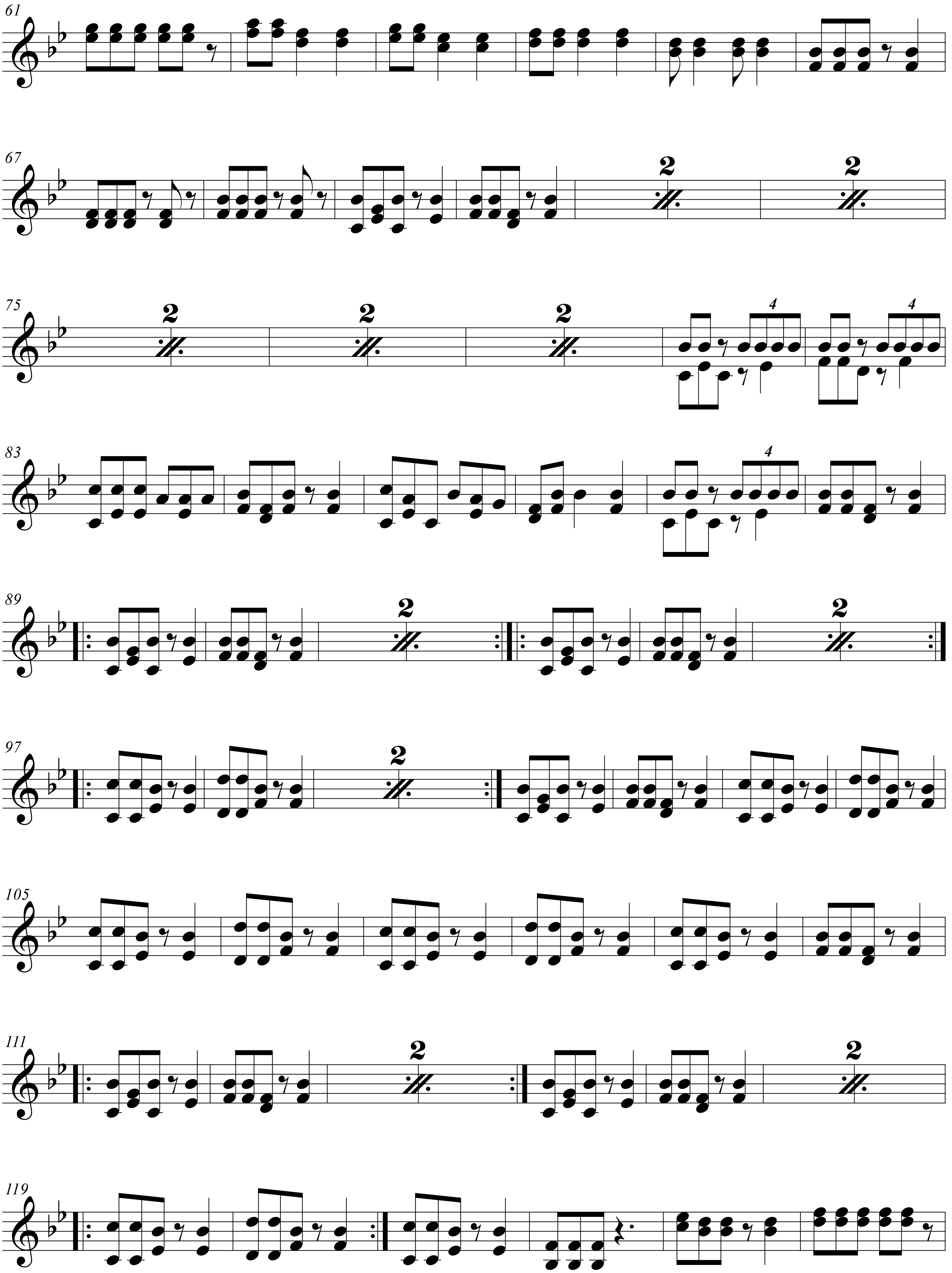

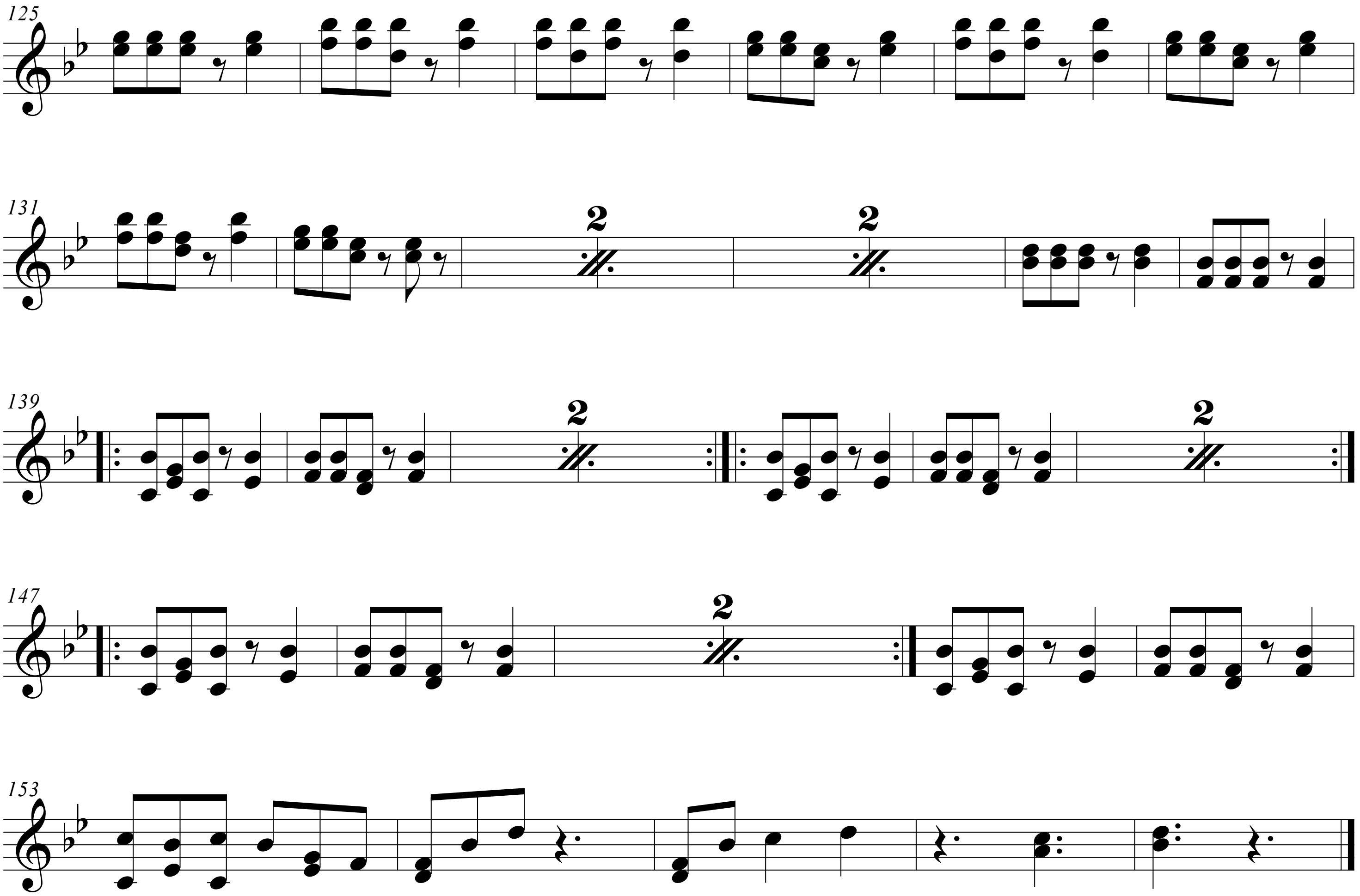

# Musical piece # 5

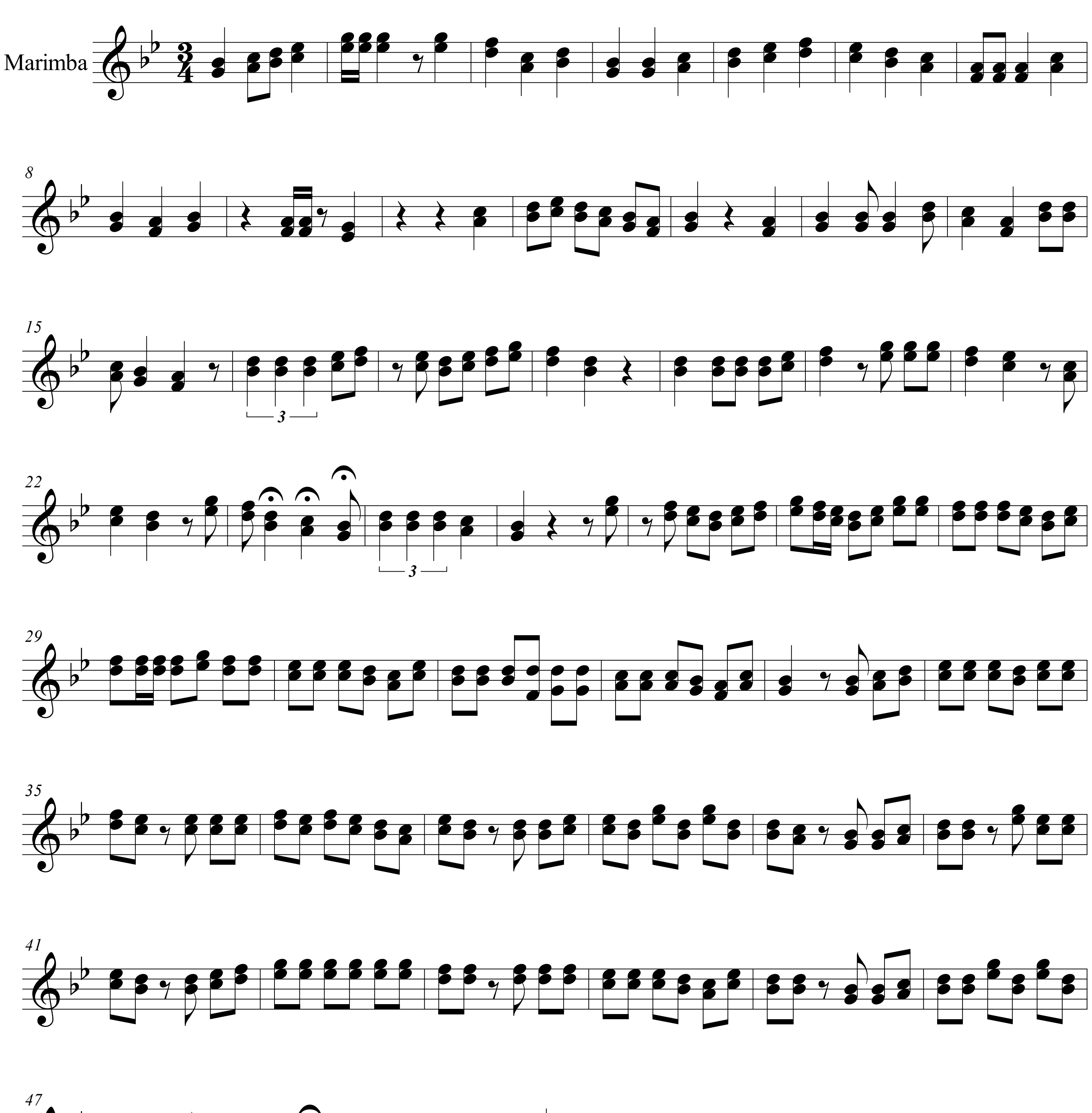

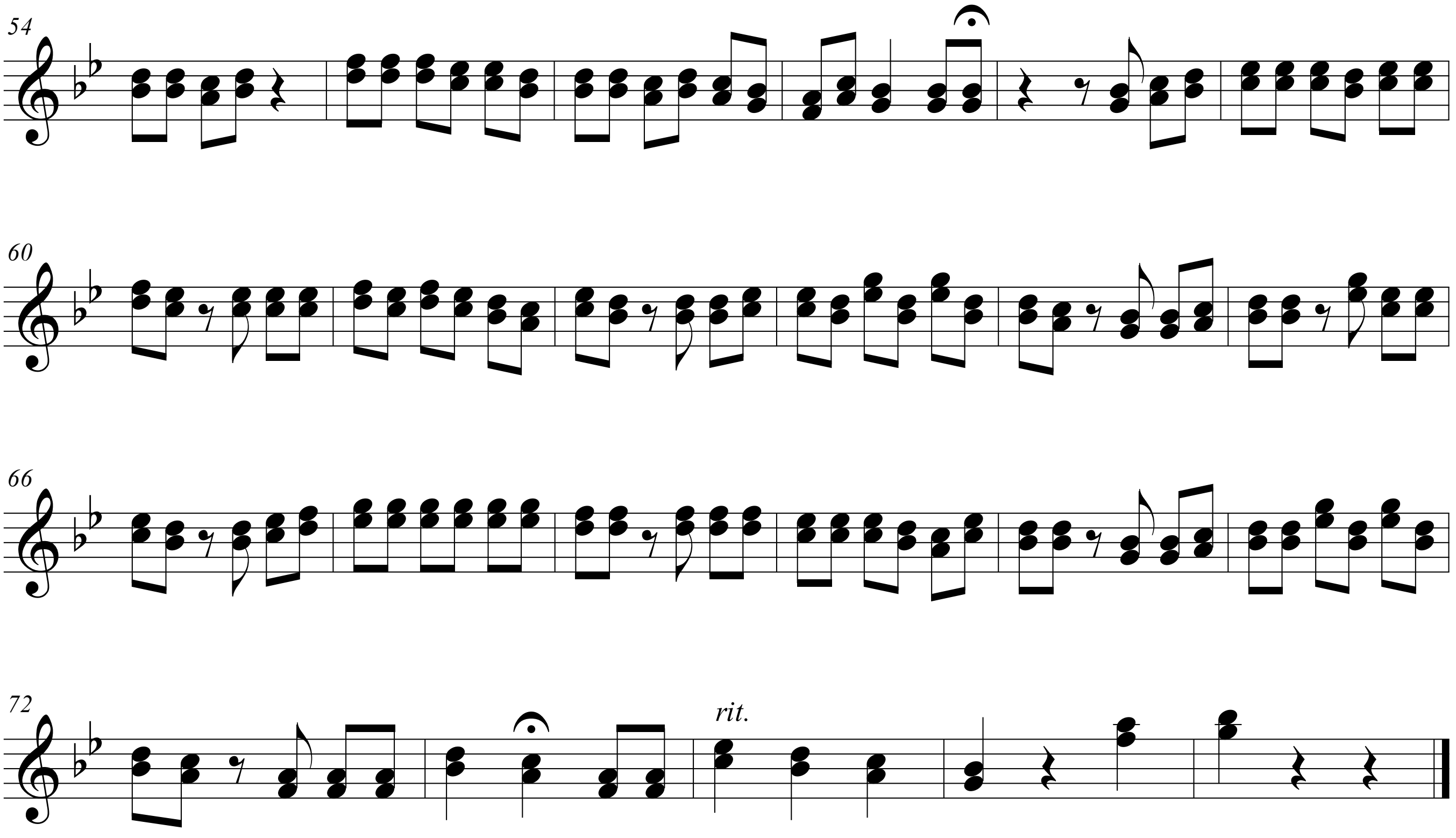
54
60
66
72
rit.

# Musical piece # 6

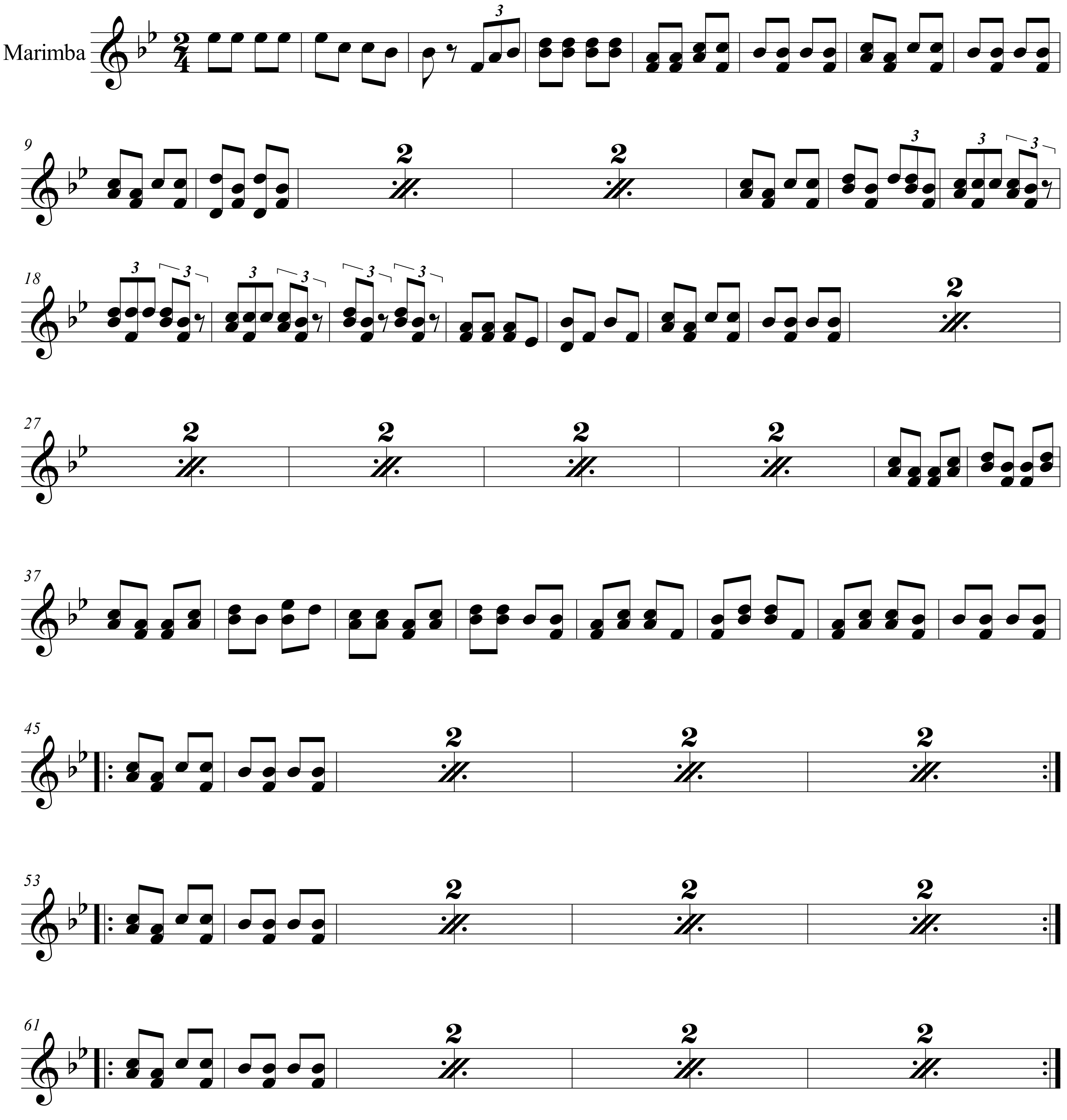

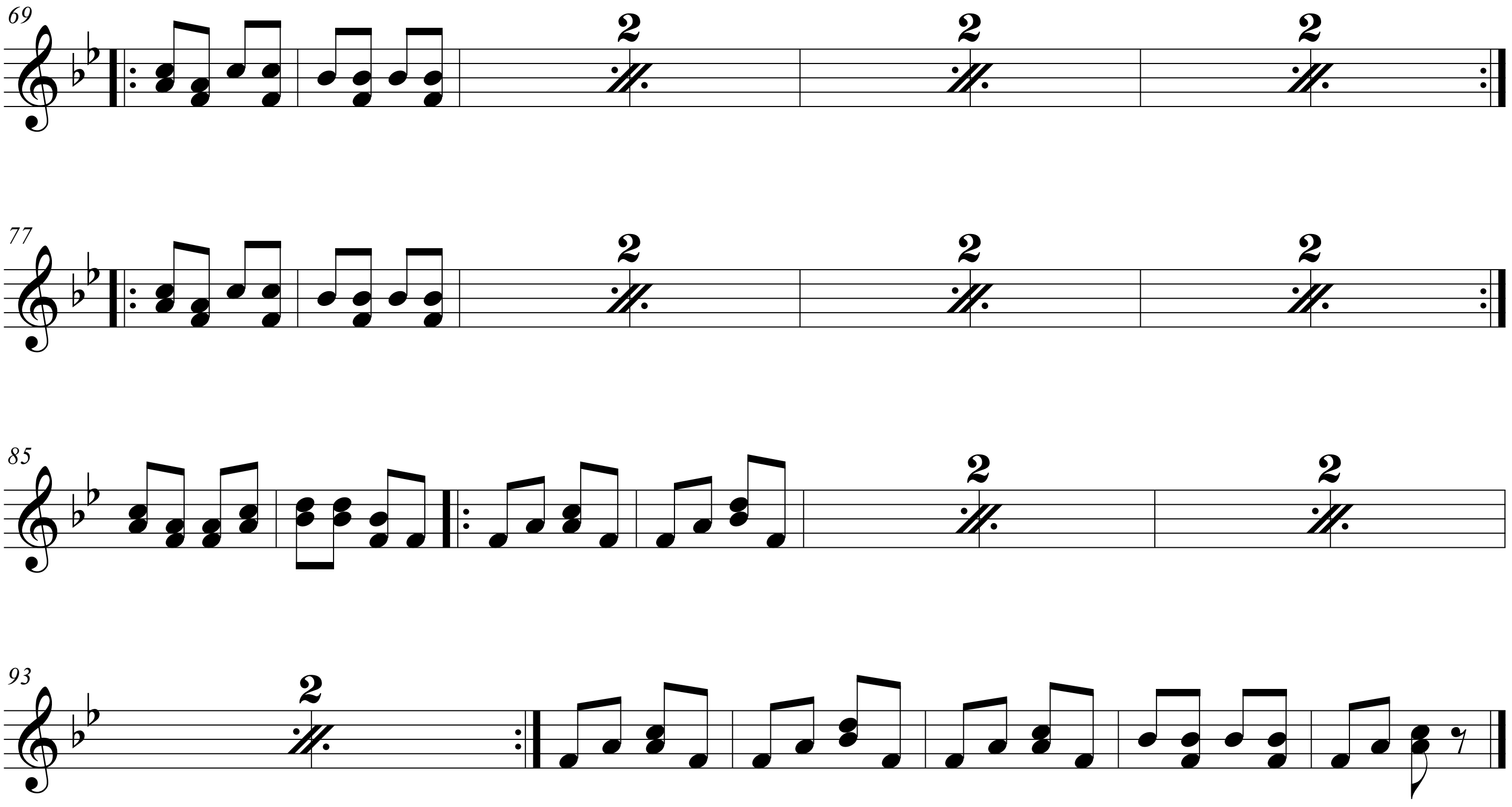

# Musical piece # 7

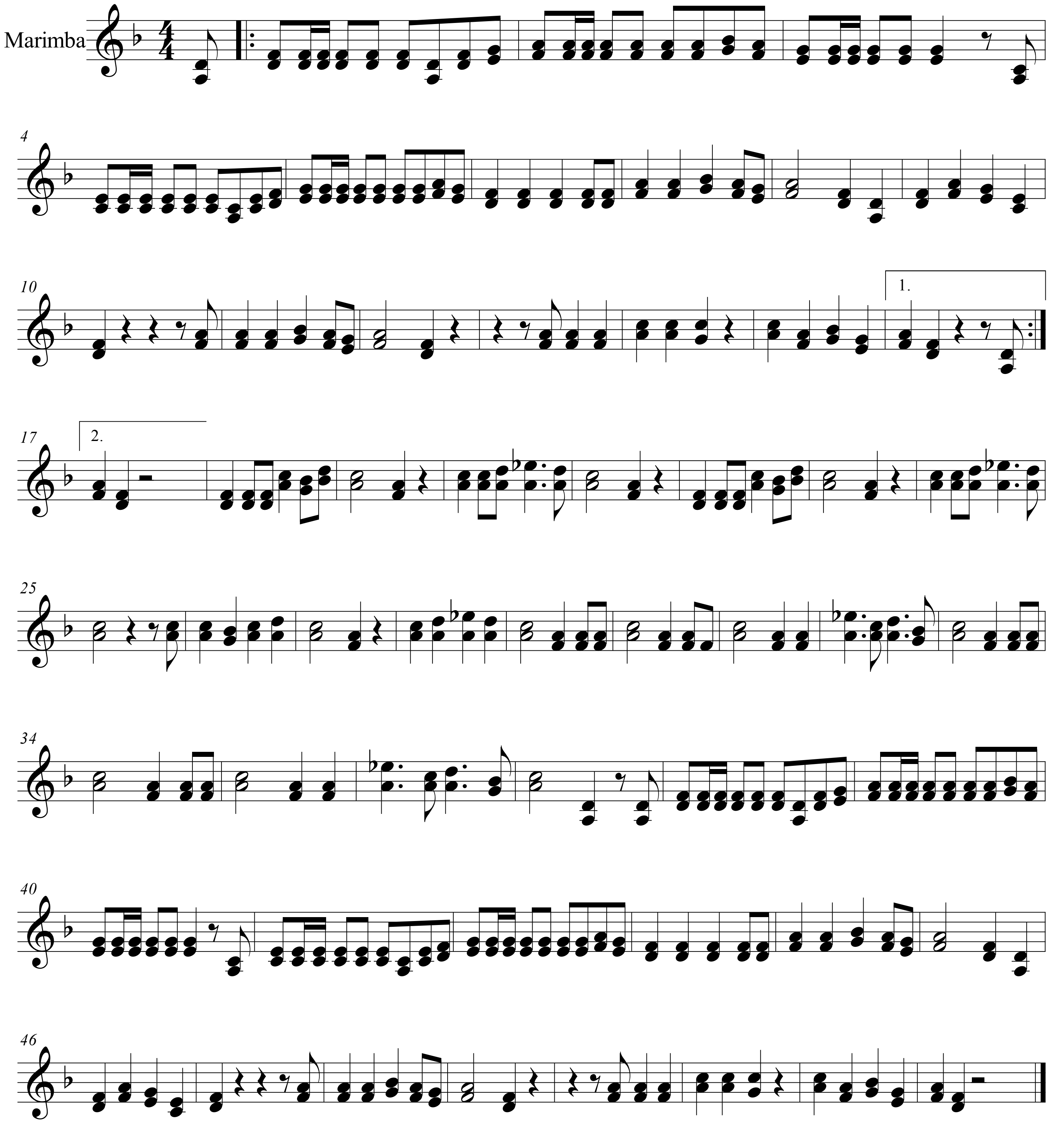

# Supplement 4

**Total number of different combinations of pairs of bars generating the same distance between them for different marimba sizes (from 14 bars up to 24 bars).**

| | | Marimba size (bars) | | | | | | | | | | |
|---|---|---|---|---|---|---|---|---|---|---|---|---|
| | Distance (steps) | **14** | **15** | **16** | **17** | **18** | **19** | **20** | **21** | **22** | **23** | **24** |
| Hexatonic | **1** | 1 | 2 | 2 | 2 | 2 | 2 | 2 | 2 | 3 | 3 | 3 |
| Hexatonic | **2** | 9 | 10 | 10 | 11 | 11 | 12 | 13 | 14 | 15 | 15 | 16 |
| Hexatonic | **3** | 4 | 5 | 5 | 6 | 7 | 7 | 7 | 7 | 8 | 8 | 9 |
| Hexatonic | **4** | 5 | 6 | 6 | 6 | 6 | 7 | 7 | 8 | 9 | 9 | 9 |
| Hexatonic | **5** | 6 | 7 | 7 | 8 | 9 | 10 | 11 | 11 | 12 | 12 | 13 |
| Hexatonic | **6** | 2 | 2 | 2 | 2 | 2 | 2 | 2 | 3 | 3 | 3 | 3 |
| Hexatonic | **7** | 6 | 7 | 7 | 8 | 9 | 10 | 11 | 12 | 13 | 13 | 14 |
| Pentatonic | **1** | 2 | 2 | 2 | 2 | 2 | 2 | 3 | 3 | 3 | 3 | 3 |
| Pentatonic | **2** | 3 | 4 | 4 | 4 | 4 | 4 | 5 | 5 | 6 | 6 | 6 |
| Pentatonic | **3** | 5 | 5 | 6 | 6 | 7 | 8 | 8 | 8 | 8 | 9 | 9 |
| Pentatonic | **4** | 4 | 5 | 6 | 6 | 6 | 6 | 7 | 7 | 8 | 9 | 9 |
| Pentatonic | **5** | 3 | 3 | 3 | 3 | 4 | 4 | 5 | 5 | 5 | 5 | 5 |
| Pentatonic | **6** | 1 | 1 | 1 | 1 | 1 | 2 | 2 | 2 | 2 | 2 | 2 |
| Pentatonic | **7** | 5 | 6 | 7 | 7 | 8 | 9 | 10 | 10 | 11 | 12 | 12 |

# Supplement 5

**Normalized amplitude for the first ten overtones produced by each bar with its corresponding resonator**. The first bar of each marimba corresponds to the lowest frequency and the last bar to the highest one.

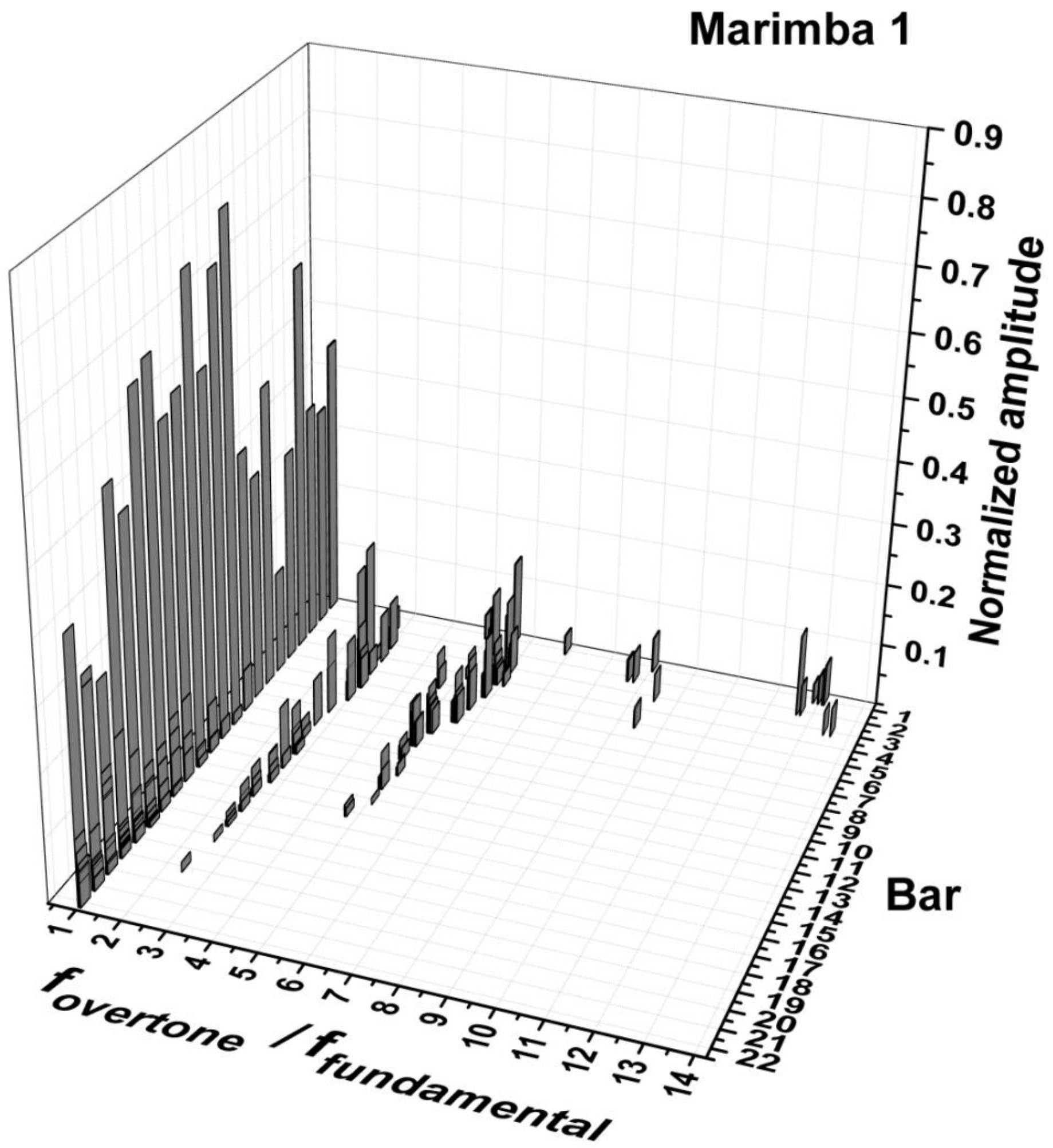

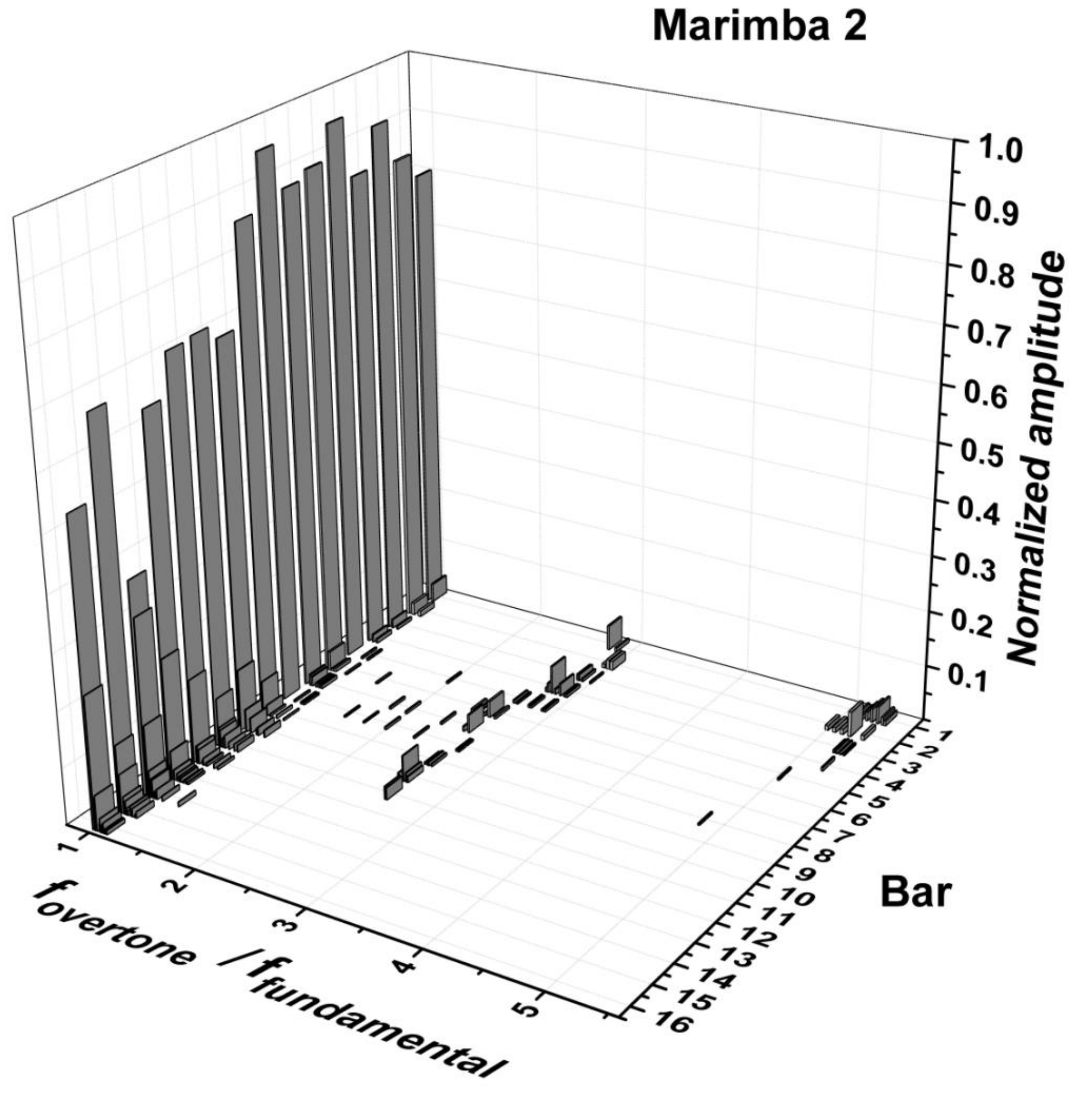

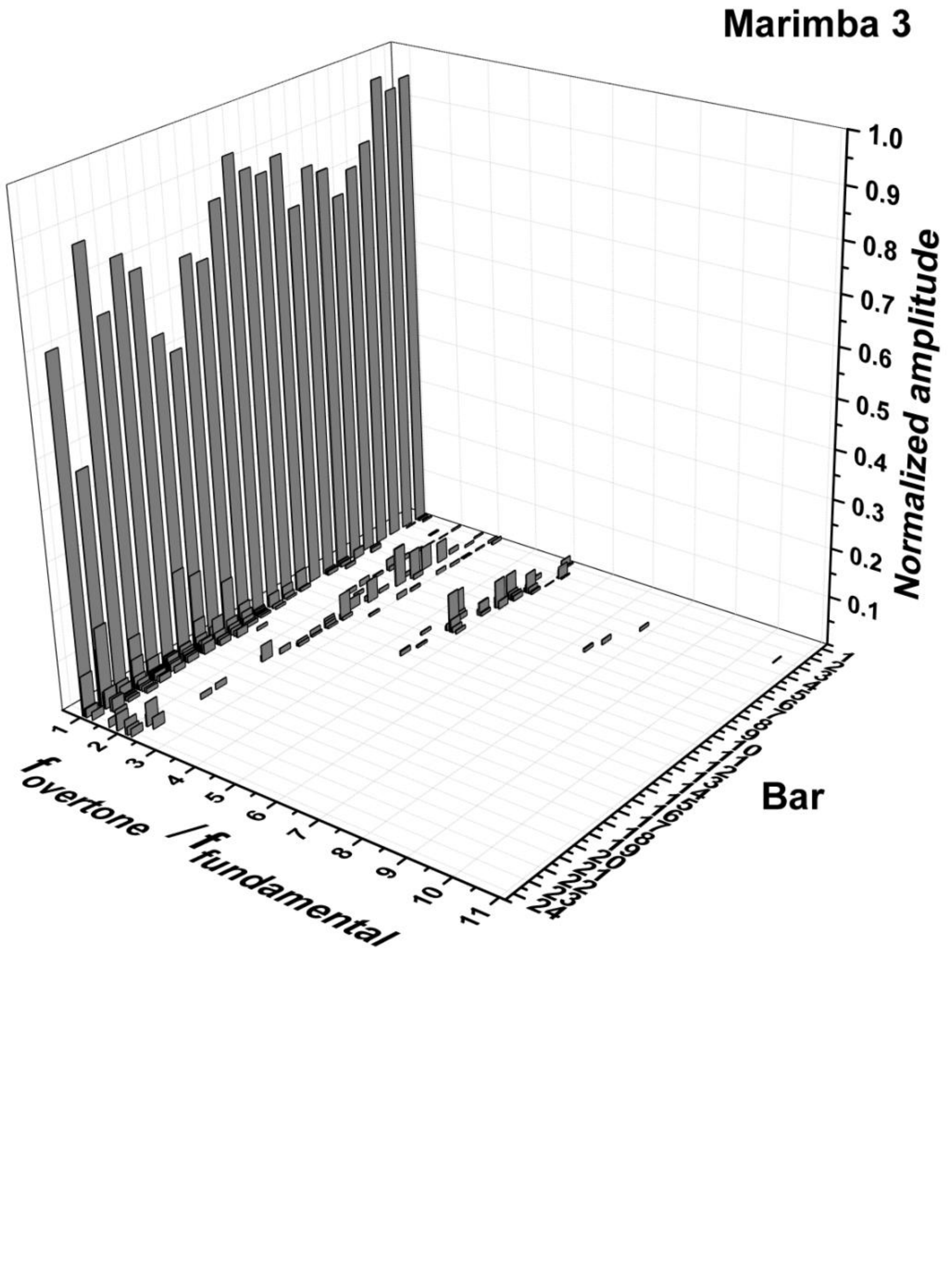

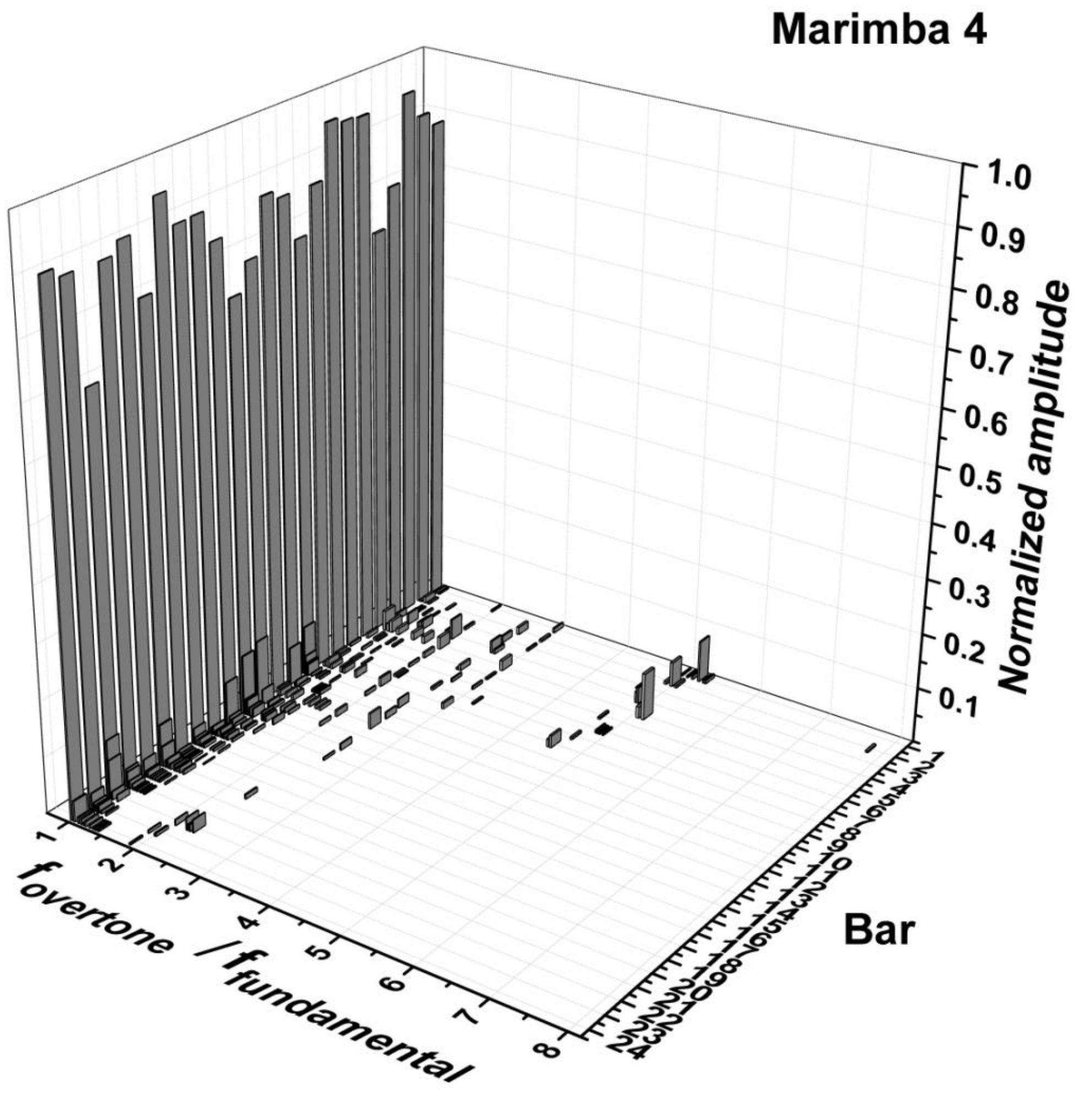

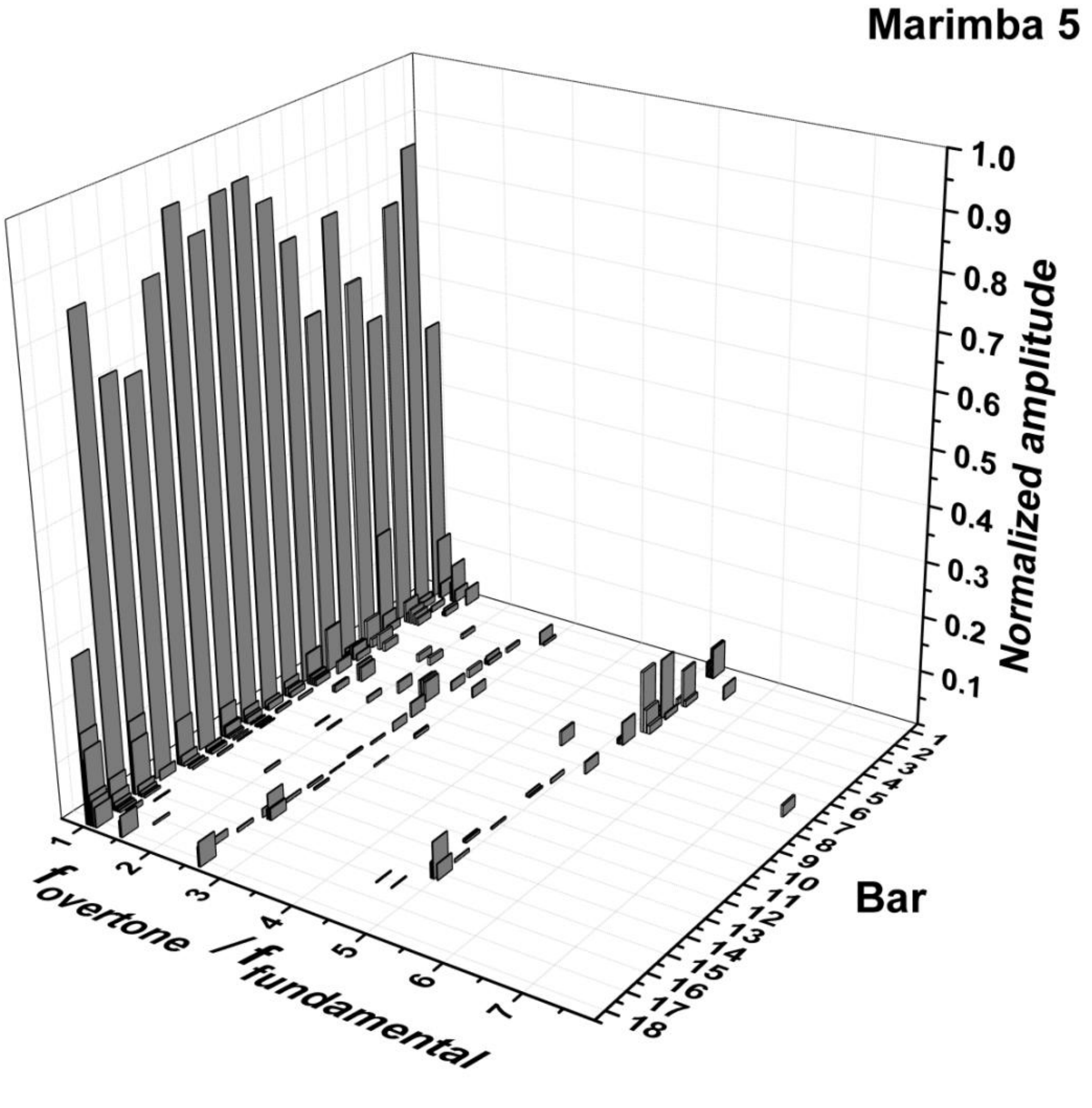

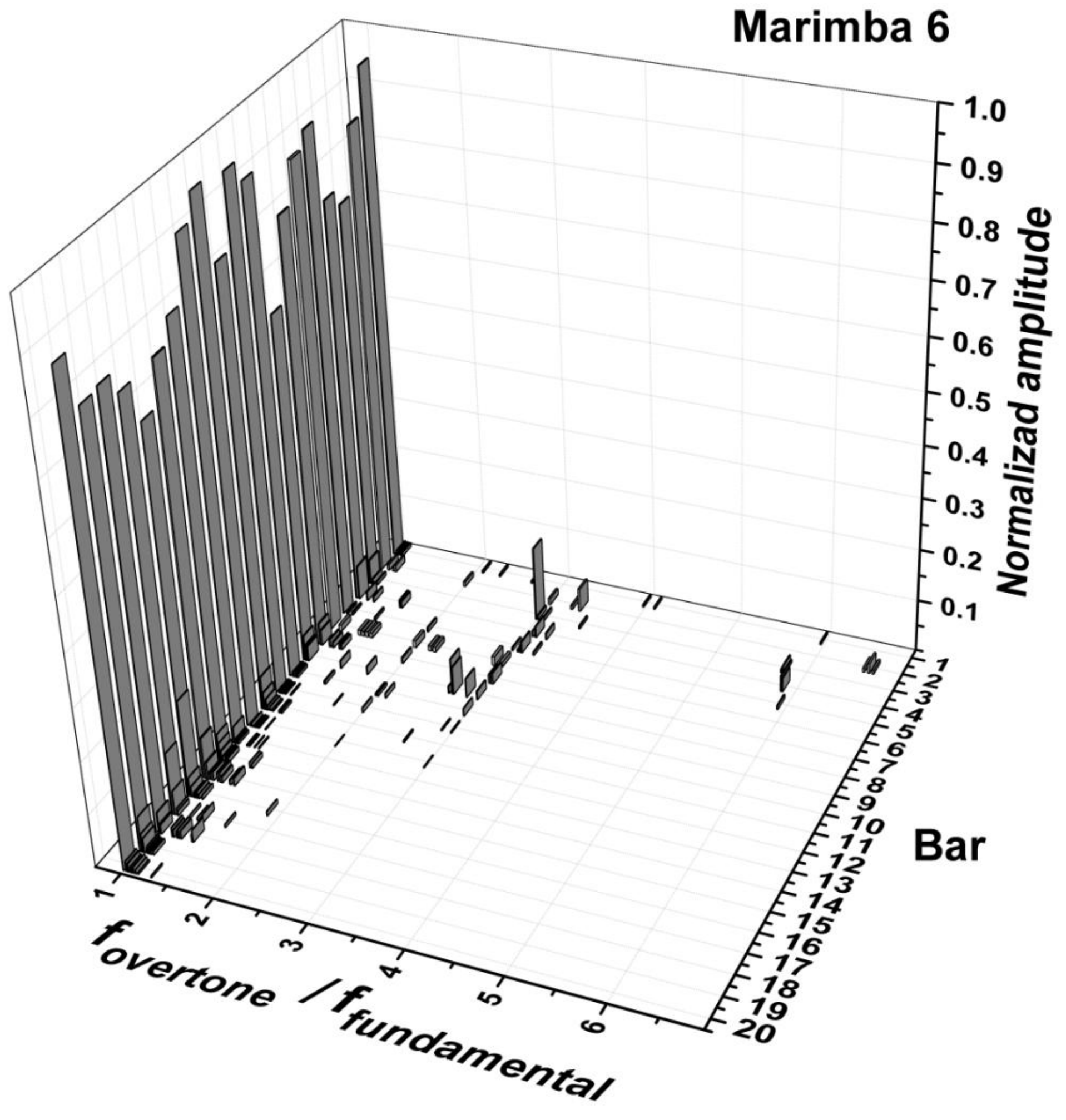

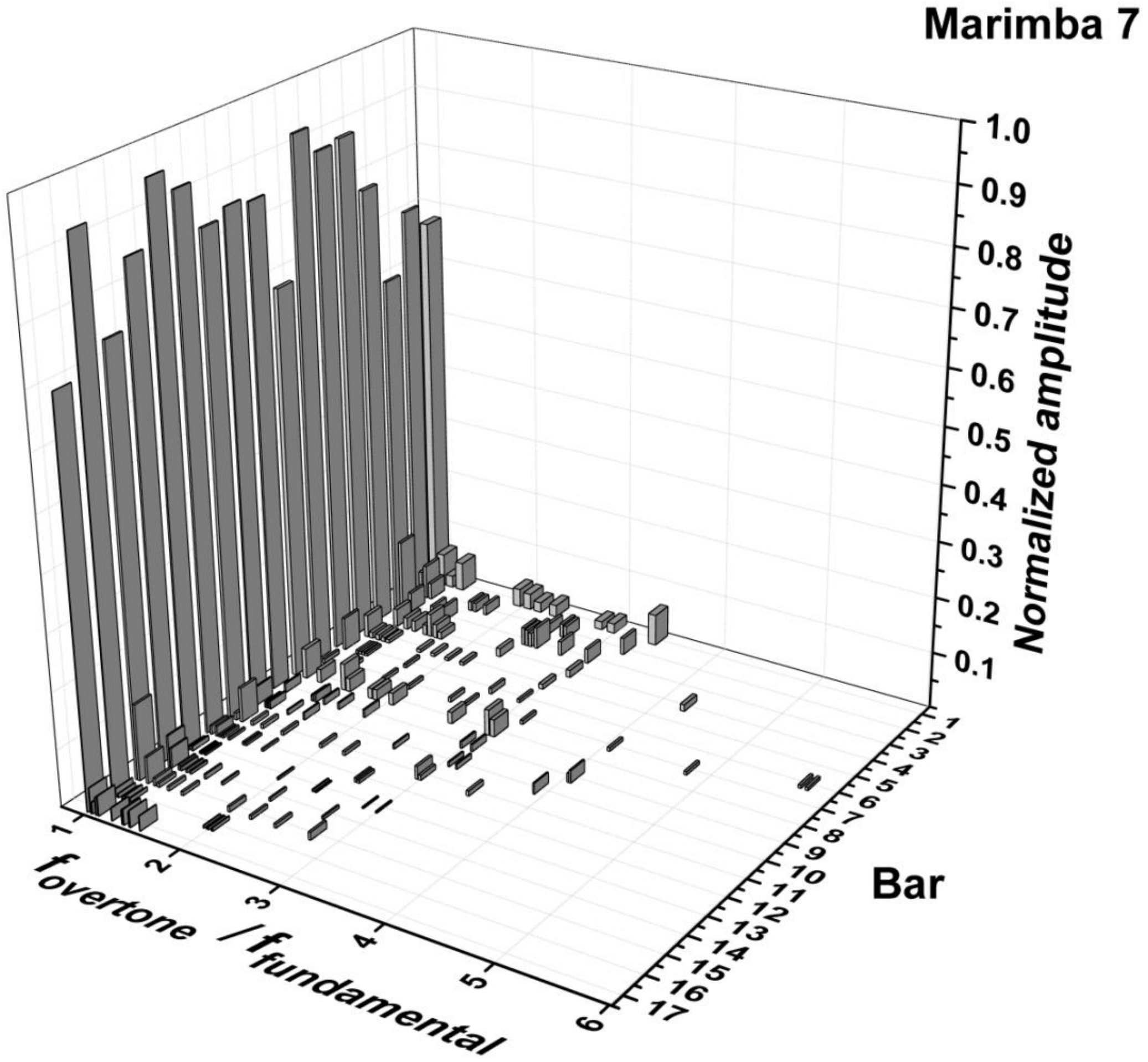

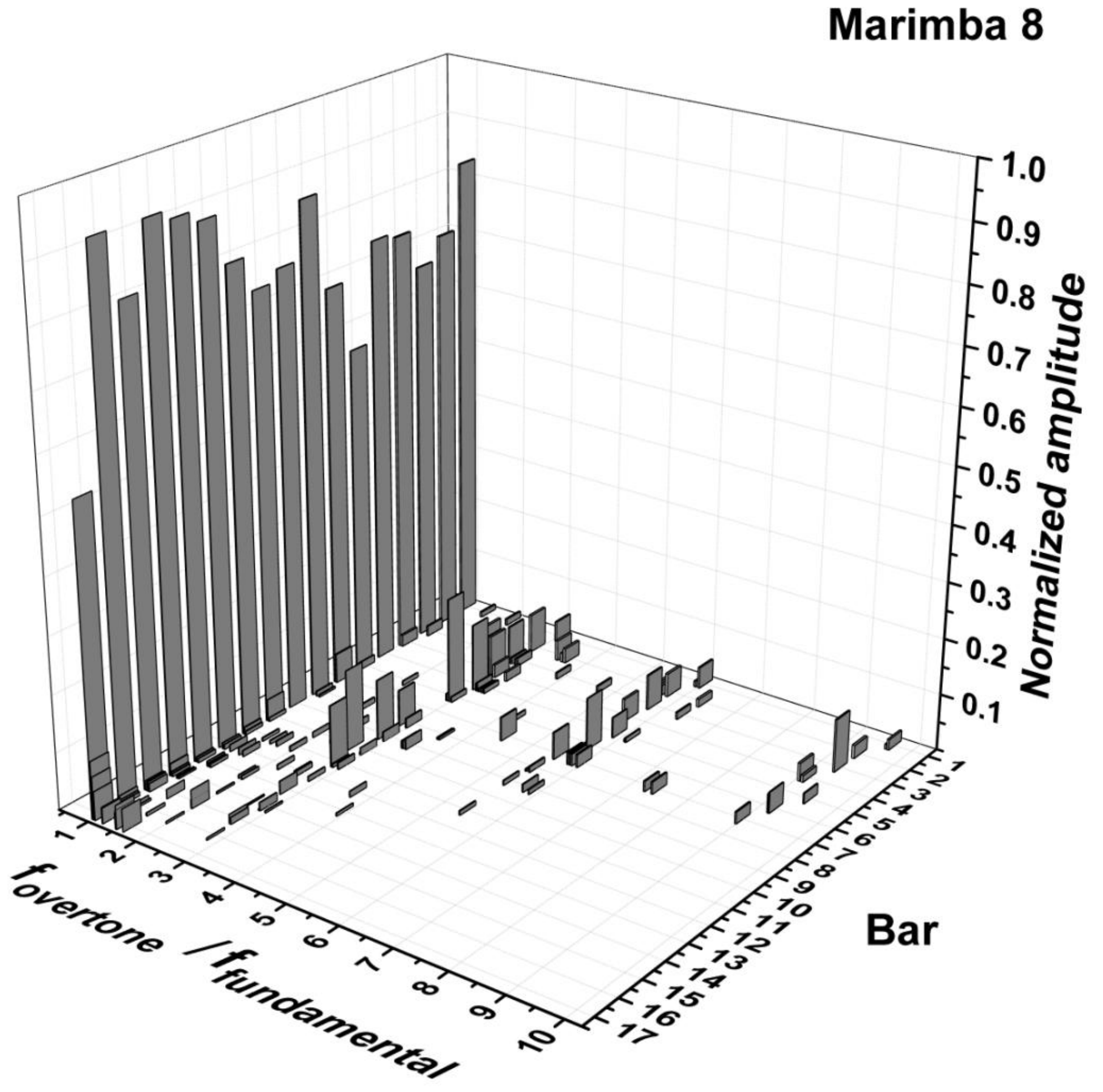

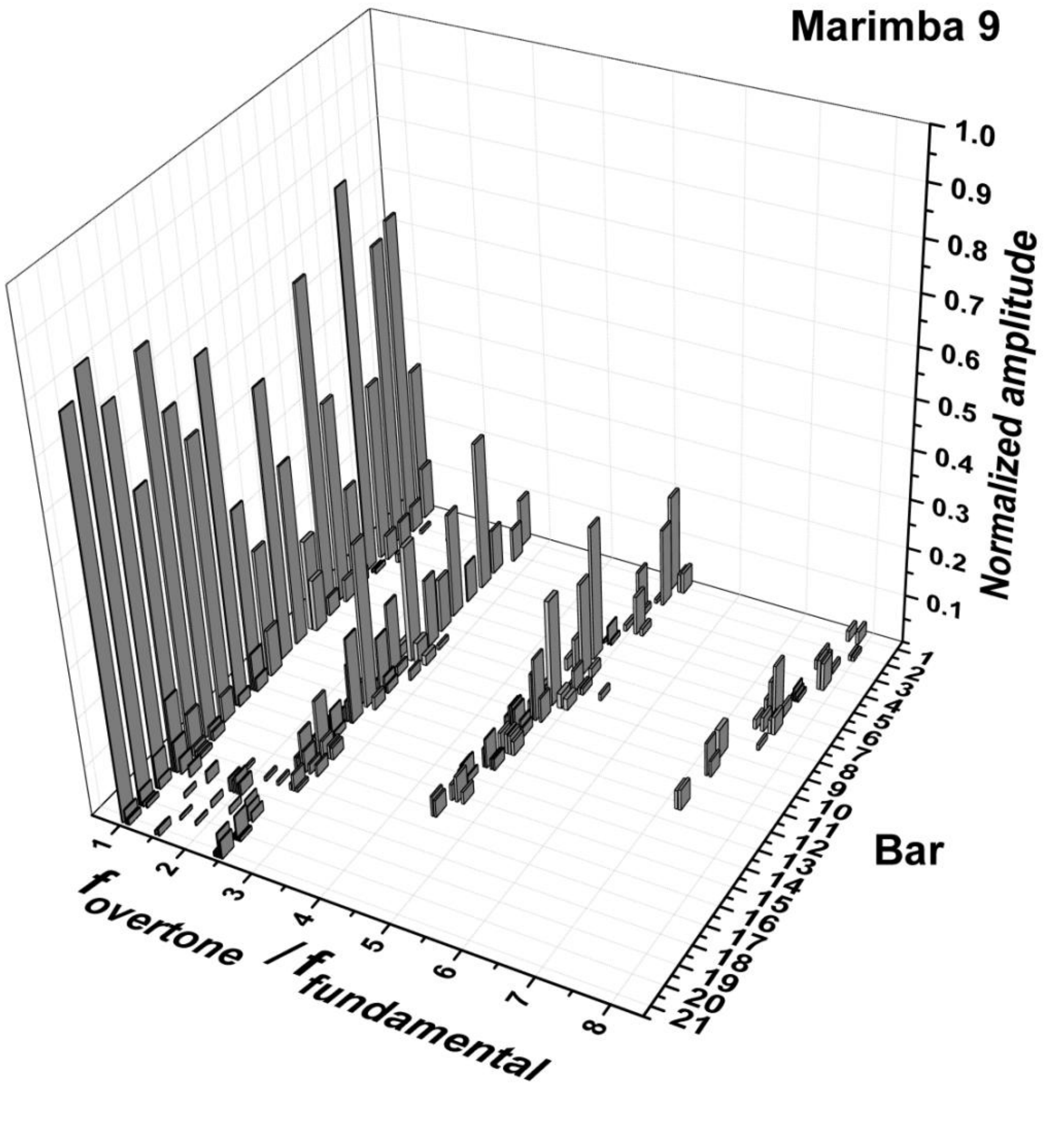

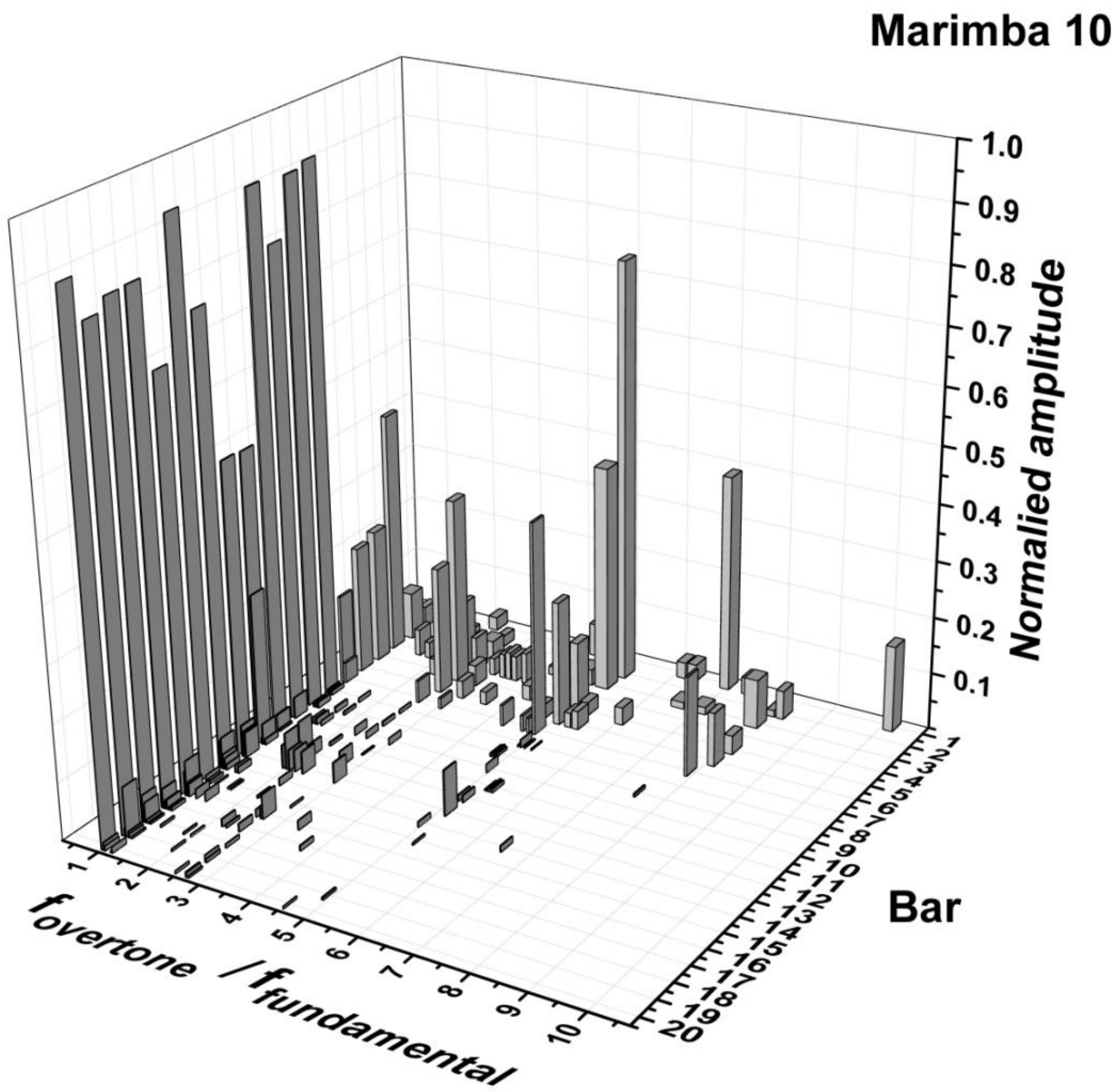

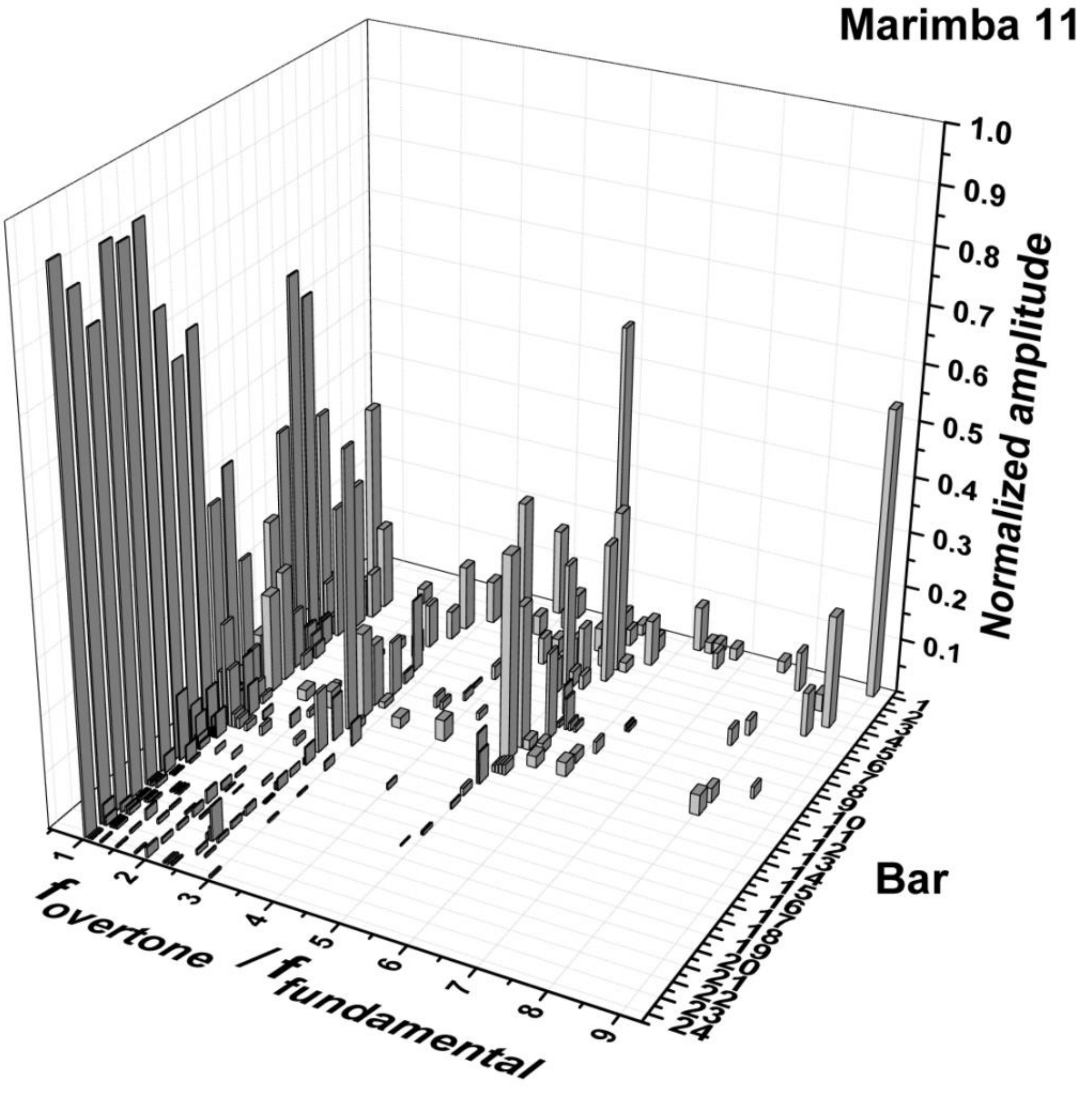